\documentclass{egpubl_arXiv}
\usepackage[T1]{fontenc}
\usepackage{dfadobe}

\usepackage{cite}  
\BibtexOrBiblatex
\electronicVersion
\PrintedOrElectronic
\ifpdf \usepackage[pdftex]{graphicx} \pdfcompresslevel=9
\else \usepackage[dvips]{graphicx} \fi

\usepackage{egweblnk}

\usepackage{amssymb}
\usepackage{latexsym}
\usepackage{amsmath}

\usepackage{url}
\usepackage{xcolor}

\usepackage{algorithm}
\usepackage{algpseudocode}
\usepackage{hyperref}
\usepackage{url}

\usepackage[switch,pagewise]{lineno} 

\usepackage{graphics}
\usepackage{graphicx}

\newtheorem{remark}{Remark}[section]

\newcommand{\eqnref}[1]{(\ref{#1})}

\definecolor{myRedColor}{RGB}{200, 0, 0}
\definecolor{myGreenColor}{RGB}{0, 125, 0}
\definecolor{myBlueColor}{RGB}{0, 0, 155}
\definecolor{myYellowColor}{RGB}{155,155, 0}
\definecolor{myCyanColor}{RGB}{0, 155, 155}
\definecolor{myMagentaColor}{RGB}{155, 0, 155}
\definecolor{myBrownColor}{RGB}{190, 100, 60}
\definecolor{myLightBrownColor}{RGB}{184, 153, 31}
\definecolor{myWhiteColor}{RGB}{255, 255, 255}
\definecolor{myBlackColor}{RGB}{0, 0, 0}

\newcommand{\SetRed}[1]{\color{myRedColor}{#1}}
\newcommand{\SetGreen}[1]{\color{myGreenColor}{#1}}

\newcommand{\SetLightBrown}[1]{\color{myLightBrownColor}{#1}}

\newcommand{\SetBlack}[1]{\color{myBlackColor}{#1}}

\newcommand{\Reals}{\mathbb{R}}

\newcommand{\Bezier}{B\'{e}zier}

\newcommand{\Bspline}{B-spline}

\newcommand{\signdist}{\text{dist}_{\pm}}

\newcommand{\MyungSoo}[1]{\SetBlack{#1}}
\newcommand{\Pablo}[1]{\SetBlack{#1}}
\newcommand{\QYoun}[1]{\SetBlack{#1}}
\newcommand{\Gershon}[1]{\SetBlack{#1}}

\title[Variable offsets and processing of implicit forms]{Variable offsets and processing of implicit forms toward the
	adaptive synthesis and analysis of heterogeneous conforming microstructure}

\author[Q. Y. Hong \& P. Antolin \& G. Elber \& M.-S. Kim]
{\parbox{\textwidth}{\centering Q.\,Y. Hong$^{1}$\orcid{0000-0002-3411-049X},
		P. Antolin$^{2}$,
		G. Elber$^{3}$,
		and M.-\,S. Kim$^{4}$
	}
	\\
	{\parbox{\textwidth}{\centering $^1$Hanyang University ERICA, Department of Computer Science and Engineering, Ansan, Korea\\
			$^2$École Polytechnique Fédérale de Lausanne, Institute of Mathematics, Lausanne, Switzerland \\
			$^3$Techion - Israel Institute of Technology, Department of Computer Science, Haifa, Israel\\
			$^4$Seoul National University, Department of Computer Science and Engieering, Seoul, Korea
		}
	}
}

\begin{document}


    \maketitle
\begin{abstract}

The synthesis of porous, lattice, or microstructure geometries has
captured the attention of many researchers in recent years.  Implicit
forms, such as triply periodic minimal surfaces (TPMS) has captured a
significant attention, recently, as tiles in lattices, partially
because implicit forms have the potential for synthesizing with ease
more complex topologies of tiles, compared to parametric forms.  In
this work, we show how variable offsets of implicit forms could be
used in lattice design as well as lattice analysis, while graded wall
and edge thicknesses could be fully controlled in the lattice and even
vary within a single tile.  As a result, (geometrically) heterogeneous
lattices could be created and adapted to follow analysis
results while maintaining continuity between adjacent tiles.  We
demonstrate this ability on several 3D models, including TPMS.


%
%
\end{abstract}

\section{Introduction}
\label{sec-intro}

The advantages of lattice based geometries in modern design are
{\Gershon becoming apparent} in more and more applications and fields.
{\MyungSoo There are applications for lattices
in many areas of advanced manufacturing}, from
lighter materials for strong enough mechanical parts, through physical
structures that mimics biological behavior in medical implants, to full
control over the heat flux through porous materials~\cite{Wang2022,Zwar2023}.

Additive manufacturing (AM) or 3D printing is a powerful enabling
technology for realizing lattices.  Graded heterogeneity and porosity are now
possible in fabrication, abilities that were unthinkable using classic
subtractive manufacturing technologies, that employ CNC machining.
With AM capabilities, one is able to design and fabricate porous
lattices and/or microstructures of great geometric complexity and even
with functional graded materials and geometries~\cite{Elber2023}.

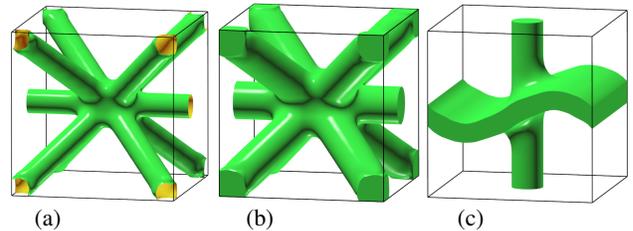
\begin{figure}
    \begin{center}
        \mbox{\hspace{-0.1in}}
        \includegraphics[width=0.32\columnwidth]{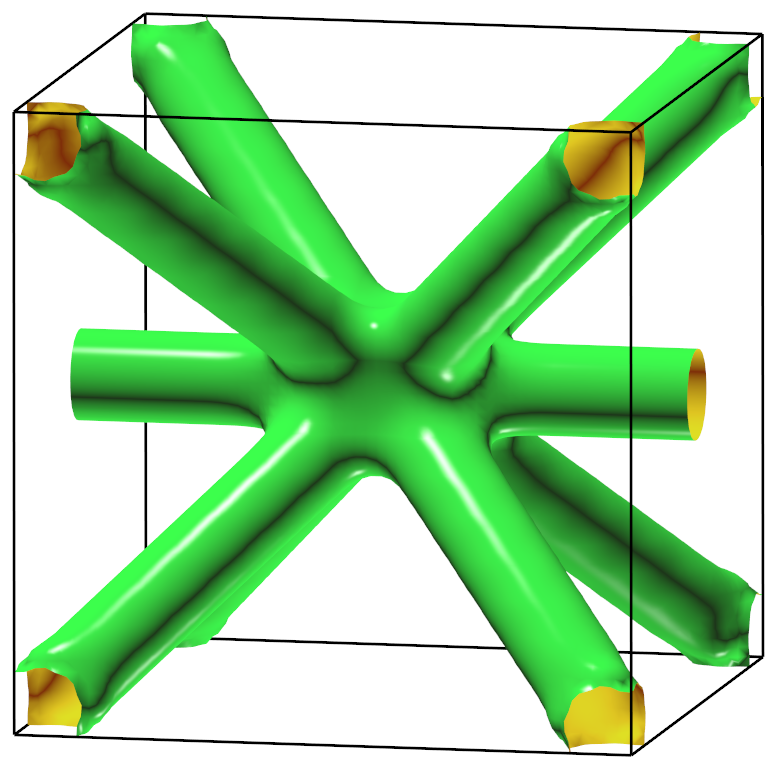}
        \includegraphics[width=0.32\columnwidth]{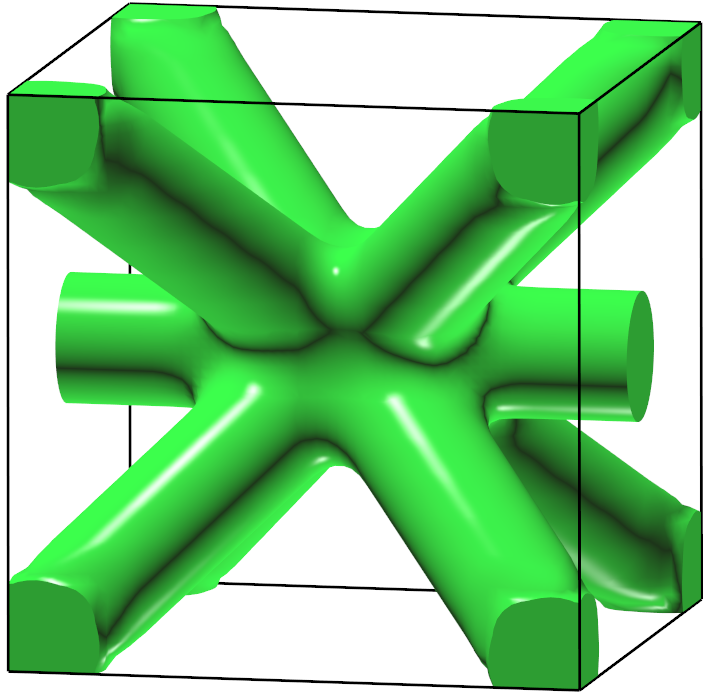}
        \includegraphics[width=0.32\columnwidth]{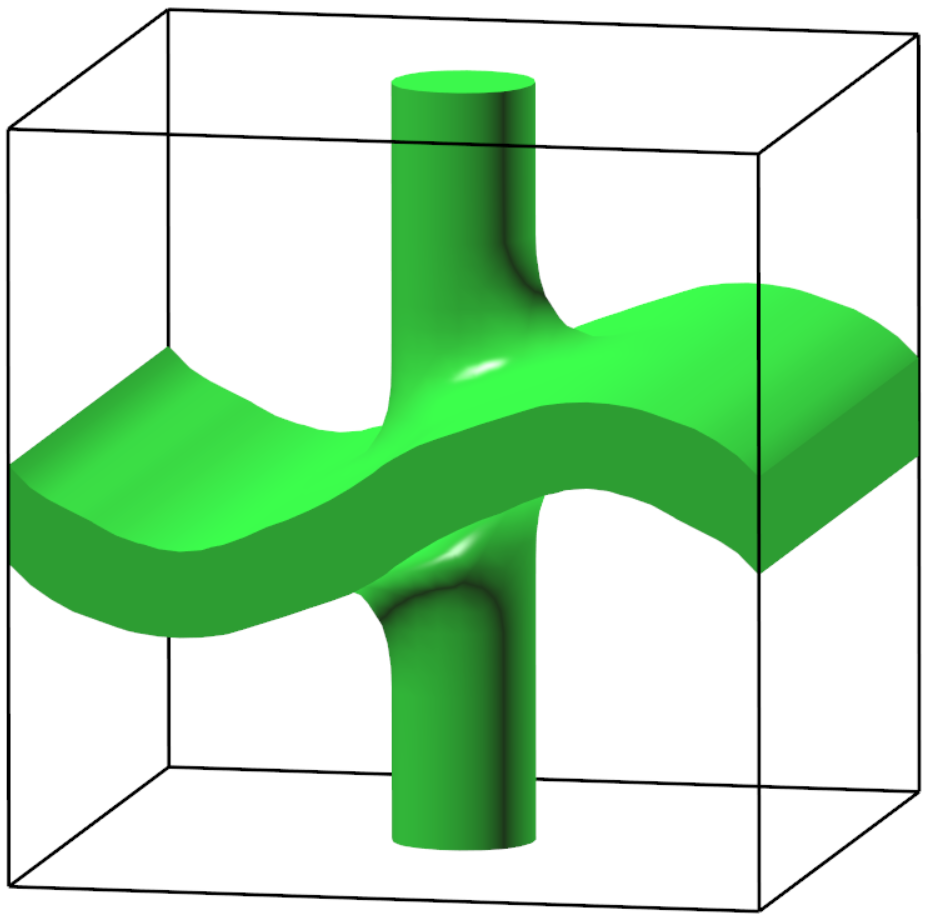}
    \end{center}
    \begin{picture}(0,0)
        \put(   10,  3){(a)}
        \put(   90,  3){(b)}
        \put(  170,  3){(c)}
    \end{picture}
    \mbox{\vspace{-0.25in}}\\[-0.25in]
    \caption{Variations of implicit trivariate \Bspline{} tiles,
        created using a distance field from curves ((a) and (b)) and
        a curve and a surface (c).  Considering the complexity of the
        central joint (with ten arms!) in (a) and (b), the recreation
        of a similar tile using parametric forms will be
        painstakingly difficult.  Note the corners are forming one
        octant of a joint with seven other neighboring
        tiles. (b) and (c) shows a closed implicit tile
        whereas in (a) the tile is created open (so it can be
        connected to some neighbors).}
    \label{fig_imp_tile_example}
    \mbox{\vspace{-0.5in}}\\[-0.5in]
\end{figure}

There are several approaches to the {\MyungSoo creation} and design of lattices,
from voxel based approaches, through spline parametric forms, to
implicit representations.  The latter are considered more versatile in
the topologies that they can represent (See
Figure~\ref{fig_imp_tile_example} for one example).  Yet, implicits
are also more difficult to manage and manipulate, in many cases,
compared to parametric forms.  For example the display of
implicits is complex as it typically requires a marching cubes-like
approximation algorithm.  Similarly, the analysis of implicit form is
a question that only recently picked some momentum, and is also
addressed here.  In Figure~\ref{fig_imp_tile_example}, a Marching
Cubes~\cite{Lorensen87} variant was employed to extract a displayable
approximated form of this implicit \Bspline{} trivariate tile.

{\Gershon The difficulties in using implicit forms compared to
parametric forms are a major reason why all modern Computer Aided
Geometric Design (CAD) systems are based almost solely on parametric
spline forms.}
Yet, the topological versatility of implicits makes them a favorable
solution in tiles' design, and in the exploitation of discrete topological
optimization results. In addition, leveraging on novel unfitted
(or immersed) finite element discretization techniques, implicits can
be directly applied as geometric descriptions for simulation purposes.
{\Gershon However, when looking for} geometric variations
within created lattices or even
within constructed tiles, wall thicknesses and edge diameters must be
controlled as well.  In this respect, one relevant fundamental
difficulty in using implicit forms is the difficulty of computing
offsets, not to say graded (variable) distance offsets.  In order to
create functionally graded geometries in implicit lattices, the wall
thicknesses and edge diameters of the implicits must vary between
tiles and even within a single tile.  This inter-tile variable wall
thickness offset as well as intra-tile variable offset is hence highly
desired and is the aim of this work.

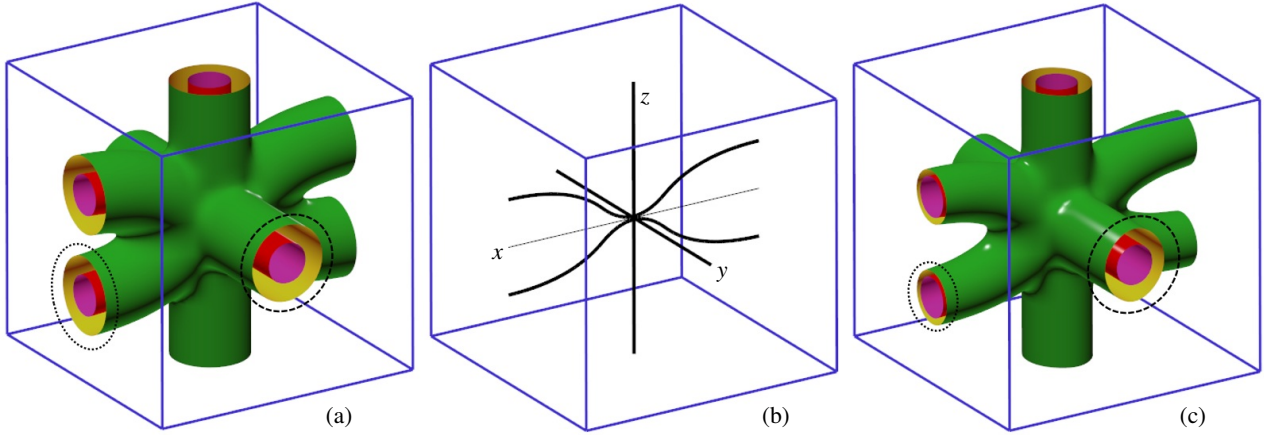
\begin{figure*}
    \begin{center}
        \includegraphics[width=0.95\textwidth]{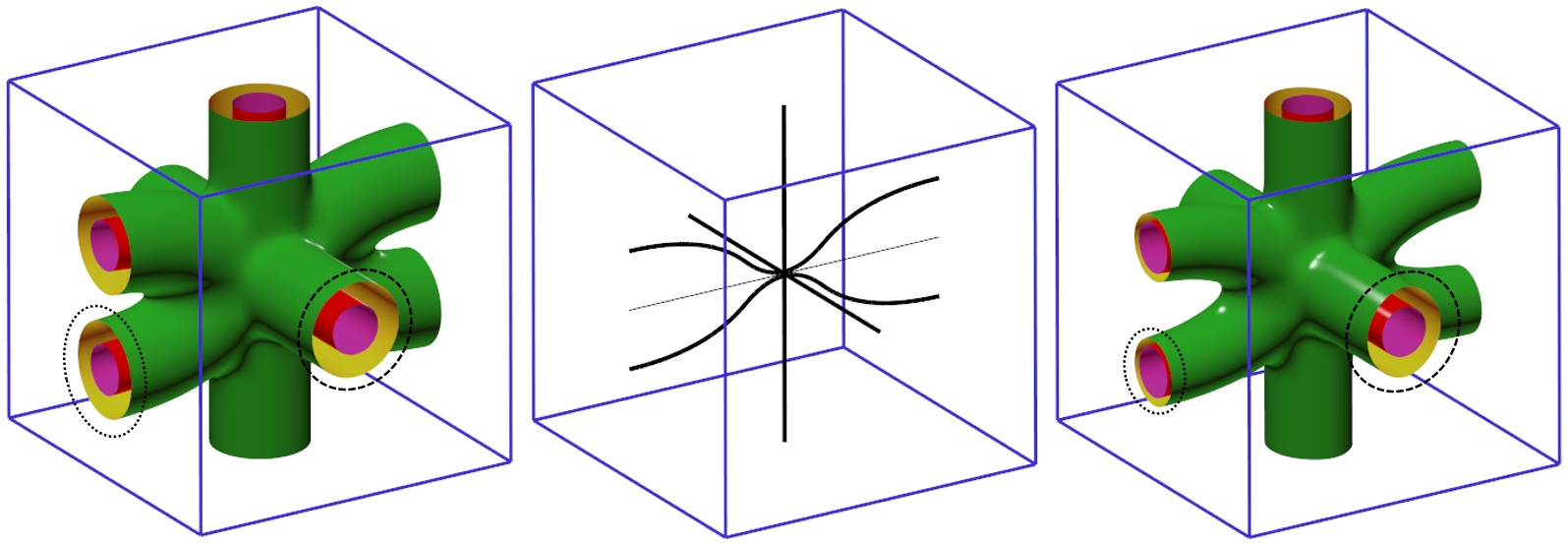}
        \begin{picture}(0,0)
            \put( -358,  5){(a)}
            \put( -193,  5){(b)}
            \put(  -35,  5){(c)}
            \put( -295,  67){$x$}
            \put( -210,  60){$y$}
            \put( -239, 126){$z$}
        \end{picture}

    \end{center}
    \caption{A counter example showing why level sets cannot be
        employed to compute precise offsets. The distance field to the
        four curves in (b) is used to derive the {\Gershon two different}
	implicit trivariates in (a) and (c), {\Gershon that are identical
	at level set $c_0$ and shown in} {\SetRed{red as ${\mathcal I}(u, v, w)
        = c_0$}} (interior in magenta).  Adjusting the level sets, in
        (a) and (c), from $c_0$ to $c_1$, yields the result in
        {\SetGreen{green as ${\mathcal I}(u, v, w) = c_1$}} (interior in
        yellow).  The {\Gershon green} level set in (c) is different from (a)
        due to graded scaled up gradients, in (c) as we move away from the
        origin, along $x$.  Note the pairs of green arms in $+x$
	{\Gershon (in the dotted ovals)} that are
        much closer to the original red arms in (c), compared to (a),
        while other arms are similar {\Gershon (e.g., in the dashed ovals)}.}
    \label{fig_level_set_vs_offset}
    \mbox{\vspace{-0.5in}}\\[-0.5in]
\end{figure*}

Given an implicit form ${\mathcal I}(u, v, w) = c_0$, one might
argue that by adjusting the constant level $c_0$, the wall thickness
of the implicit form can be controlled.  Unfortunately, and due to the
fact that the magnitude of the gradients of implicit forms are rarely
constant, not to say unit size, level set values cannot be employed to
control precise offsets.  In general, there is no simple global
relation between the value of $c_0$ and an offset amount or desired
wall-thickness, as the (magnitudes of the) gradients can vary
arbitrarily across the domain.  See Figure~\ref{fig_level_set_vs_offset}
for a counter example.

By solving the eikonal equation: $|{\nabla {\mathcal I}(u, v, w)}| = 1$,
one may aim to approximate the offset surface in the form of
${\mathcal I}(u, v, w) = c_0 + d$, which requires a numerical integration
over the region bounded by two surfaces:
${\mathcal I}(u, v, w) = c_0$ and ${\mathcal I}(u, v, w) = c_0 + d$~\cite{Zhao2005}.
However, the offset surface is defined more efficiently using only
the given surface ${\mathcal I}(u, v, w) = c_0$  and its normal field ${\nabla {\mathcal I}(u, v, w)}$
of the surface. In fact, the offset surface can be approximated precisely
by sampling the distances to the surface from the neighborhood of
the offset surface. We employ the B-spline multivariate function
which is very useful for this purpose due to the local shape control properties.

The rest of this work is organized as follows. In
Section~\ref{sec-prev-work}, we discuss relevant previous work, mostly
on implicit tiles in lattice design.  Section~\ref{sec-algorithms}
portrays the different algorithms we employ in building implicit
\Bspline{} forms.  In
Section~\ref{subsec-create-implicit-tiles}, we present simple ways to
build implicit \Bspline{} trivariatres, for instance, using distance
fields and in Section~\ref{subsec-create-implicit-offset} we show how
to compute constant and variable offsets of implicits.  However,
herein the implicit tiles are mapped through a trivariate form
${\mathcal T}(u, v, w)$, creating a lattice in the shape of ${\mathcal
T}$, in Euclidean {\em lattice space}.  As a result, constant (and
variable) offsets of implicit tiles in the domain of ${\mathcal T}$
will no longer be preserved, after the mapping through ${\mathcal T}$,
in its {\em lattice space} range.  In
Section~\ref{subsec-prescribed-offsets}, we hence explore how to
compute constant and variable offsets in the range of ${\mathcal T}$,
in the Euclidean lattice space.  In
Section~\ref{sec-analysis-implicit}, we discuss (along
with {\MyungSoo Appendix}~\ref{sec-appendix}) an immersive approach toward the finite
element simulation on implicitly defined domains and lattices,
closing the loop between design and analysis.  Then, in
Section~\ref{sec-results}, some results are demonstrated, including of
analyses of the synthesized geometries.  Finally, in
Section~\ref{sec-conclude}, we conclude.


\section{Previous Work}
\label{sec-prev-work}

Additive manufacturing introduces new technologies for modeling and
fabricating heterogeneous volumetric models with microstructures~\cite{Yan2019}.
Triply periodic minimal surfaces (TPMS) are based on implicit
representations for microstructures populated within repetitive cuboid
volume elements.  The thickness of these microstructures is often
controlled by the level set threshold parameters (that we
just showed do not correspond to offsets) for their implicit
representations~\cite{Hu2021,Feng2021,Gao2022,Gao2024,Lehder2021,Maskery2018,Yan2023}.
Hu and Lin~\cite{Hu2021} proposed an approach that represents the threshold
as a trivariate \Bspline{} function, called \textit{a threshold
distribution field (TDF)}.
Using porous synthesis and implicit
\Bspline{} functions, Gao et al.~\cite{Gao2022,Gao2024} approximated porous
structures more general than those based on TPMS models.
This reliance on the threshold level has little geometric meaning and
it is difficult to (locally) control the thicknesses using it.
As an alternative, by employing geometric operations such as minimum
distance and offset computations, one can take a systematic approach
that is based on the Euclidean distance field from the microstructure
surfaces, which can be interpreted as variable-radius offsets from the
implicit surfaces that bound the microstructures.

Implicit modeling is a method of choice for representing complex 3D
shapes with non-trivial topologies and branching structures such as
trees and blood vessels~\cite{Blinn1982,Bloomenthal1995,Pasko2011,Wyvill1999,Zanni2013}.
This is demonstrated, for example, in the commercial software of
nTopology that provides handy and robust tools for modeling and
fabricating microstructures in many industrial and biomedical
applications in practice~\cite{nTopology}.
At the core of the new
approach of implicit modeling is the power of computing Euclidean
distance fields in a highly reliably way, often with the hardware
support of modern GPUs.
Using distance fields
sampled at high resolutions, Boolean and offset operations become
considerably easier and more robust to implement than other
conventional methods.  Nevertheless, the resolutions are yet limited
and it is still quite cumbersome to support local modifications to the
objects under a shape design process.


In this work, and in an effort to deal with the issue of discrete
resolutions, we employ implicit \Bspline{} functions in a more general
modeling environment such as the functional mapping of implicit
models~\cite{Hong2021,Hong2023,Massarwi2016,Massarwi2018,Massarwi2019}.
Though yet an approximation method for the (variable-radius) offset
surface construction, the problem itself poses multi-sided technical
challenges due to the underlying multi-step non-linear representations
as the result of composition of \Bspline{} functions and constraints,
etc.

Efficient and numerically stable computation of offsets for curves and
surfaces play an important role in many geometric operations (such as
rounding, filleting, and shelling) for solid
modeling~\cite{Barnhill,Hoschek93,Rossignac1986}.  Conventional offset
algorithms mostly deal with parametric curves and surfaces that form
the boundary of solid objects~\cite{Cohen2001,Hoschek93}.
Unfortunately, the offsets of rational curves and surfaces are
algebraic, but not rational in general, except for some special
cases~\cite{FaroukiBook}.  This fundamental limitation has motivated
the development of a large body of alternative methods that
approximate offset curves and surfaces in rational forms.
There have
also been a few previous methods that represent the exact offsets as
algebraic curves and surfaces (which are implicitly defined by
polynomial equations)~\cite{Hoffmann1989}.  The main computational
bottleneck in this direct approach is in the algebraic degrees of the
exact offsets, which are considerably higher than the given curves and
surfaces.  Due to the high degrees of the exact offsets, many
redundant components are often generated, which are extremely
difficult to deal with, in the offset trimming procedure.

The exact offsets of implicit polynomial curves and surfaces are also
high-degree algebraic curves and surfaces with many redundant
components. For non-polynomial implicit surfaces such as triply
periodic minimal surfaces (TPMS) defined with trigonometric functions,
the exact offsets are even more difficult to compute.  In the current
work, we consider the problem of approximating offsets of surfaces
procedurally defined with functional compositions over implicitly
defined surfaces in highly general forms.  A typical example of such
surfaces is a trivariate volume populated with repetitive tiles
defined by implicit TPMS surfaces under non-linear
freeform deformation~\cite{Hong2023}.  To give desired
uniform/non-uniform thickness, optionally with local
control, to these microstructure surfaces, we develop effective
offset approximation methods for both constant and variable-radius
offsets.

There is a long history of developing offset approximation
methods~\cite{Elber97,Maekawa99}.  However, it is yet highly
non-trivial to deal with the offset trimming of removing all redundant
components from the offset curve segments and surface patches thus
approximated~\cite{Hong2019}.  Though with somewhat limited precision
of approximation, a highly effective approach for robust
implementation of offset computation can be based on the computation
of distance fields, such as the methods proposed by
Kobbelt~\cite{Kobbelt}, Wang and Manocha~\cite{CLWang}, and Li and
McMains~\cite{McMains2014}.  We take a similar approach but represent
the approximated distance field in a \Bspline{} form, where the
constant and variable-radius offset surfaces can be extracted by
marching over the \Bspline{} distance function defined on a trivariate
volume in the Euclidean space.  The representation in \Bspline{} forms
makes it easy to overcome the limited resolution of the voxelized
space for sampling the distances.
More importantly, the basic framework is not only limited to the Euclidean
distance, and it is applicable to diverse application environments
for the synthesis and analysis of heterogeneous microstructures.

Regarding simulation of lattice geometries defined through
implicits, the current main limitation is the high computational cost
of classical finite element discretization techniques, that are greedy
in terms of computational resources.  Thus, most of the available
results, including some commercial solutions, are limited to either
shell (2D) models~\cite{Wang2020} or a small number of
tiles~\cite{Abueidda2019}; or they apply model reduction techniques
such as multiscale or numerical homogenization~\cite{Kochmann2019},
both relying on the separation of scales, what it is usually not true
for most of the applications due to the achievable length-scale of
current 3D printers.  Worth mentioning exceptions are the
works~\cite{Korshunova2021a,Korshunova2021b} in which the authors
leveraged on immersed finite element techniques, similar to the ones
applied in this work, for studying the behavior of printed lattice
designs using images, or the Immersed Method of Moments developed by
Intact Solutions and applied to the simulation of implicitly defined
lattices~\cite{IntactSolutions}.


\section{Algorithms}
\label{sec-algorithms}

\begin{figure}
    \begin{center}
        \includegraphics[width=0.9\columnwidth]{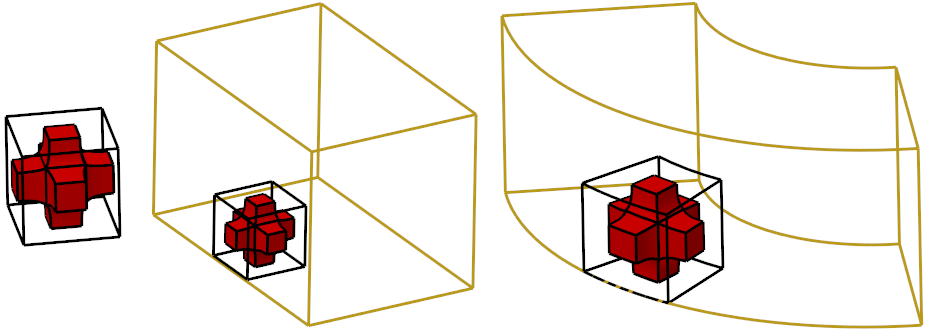}
        \begin{picture}(0,0)
            \put( -221,  12){${\mathbf I}(u, v, w)$}
            \put( -131,  70){\SetLightBrown \large $D$}
            \put( -179,  40){\footnotesize ${\mathbf I}(u, v, w)$}
            \put( -170,  16){\vector(4,-3){10}}
            \put( -178,   5){\tiny $u \approx t_x$}
            \put( -158,  10){\vector(4, 1){13}}
            \put( -143,  10){\tiny $v \approx t_y$}
            \put( -171,  20){\vector(0, 1){13}}
            \put( -182,  34){\tiny $w \approx t_z$}
            \put(  -48,  10){${\mathcal T}({\mathbf I})$}
            \put(  -45,  70){\SetLightBrown \large ${\mathcal T}(D)$}
        \end{picture}
    \end{center}
    \mbox{\vspace{-0.4in}}\\[-0.4in]
    \caption{The relationship between an implicit trivariate tile
        ${\mathbf I}(u, v, w)$ and the parametric trivariate
        macro-shape ${\mathcal T}(t_x, t_y, t_z)$.  Note that, for
        every tile ${\mathbf I}$ in ${\mathcal T}$, the tiles' $(u, v,
        w)$ parametric values affinely equates (denoted by $\approx$)
        with the parameters of a box sub-domain of the domain $D$ of
        ${\mathcal T}$, or $(u, v, w) \approx (t_x, t_y, t_z)$, for
        every tile in $D$.}
   \label{fig_implicit_tiling}
   \mbox{\vspace{-0.5in}}\\[-0.5in]
\end{figure}
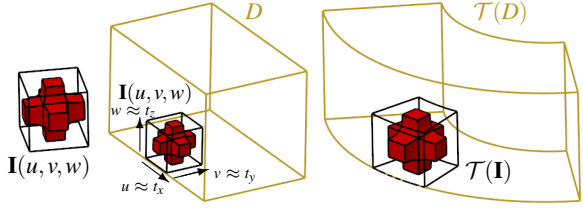

Consider a parametric trivariate ${\mathcal T}(t_x, t_y, t_z): D \rightarrow
\Reals^l$, $l \ge 3$ where the 3D geometry of ${\mathcal T}$ is defined
for $1 \le l \le 3$ and an arbitrary set of material properties are
optionally prescribed for $l \ge 4$.  ${\mathcal T}$ will be denoted the
\textit{macro-shape} hence after.  One can arrange a 3D grid of
implicit tiles, ${\mathbf I}_{ijk}(u, v, w)$, $u, v, w \in [0,1]$ in
domain $D$, so $(u, v, w)$ of each tile affinely (up to
translation and scale) equates with $(t_x, t_y, t_z)$, for some box
sub-domain of $D$, (See Figure~\ref{fig_implicit_tiling}), only to
evaluate them through ${\mathcal T}$, yielding a lattice in the shape of
${\mathcal T}$.  One should note that while the direct composition of
${\mathbf I}_{ijk}$ and ${\mathcal T}$ is difficult, point evaluation is
simple and will be denoted as ${\mathcal T}({\mathbf I}_{ijk})$.

Now, consider one such implicit tile ${\mathbf I} \subset D$.  ${\mathcal
    T}({\mathbf I})$ can inherit from ${\mathcal T}$ certain properties, such as
RGB colors, and can also inherit physical properties such as local
wall thicknesses and/or local diameters of arms (recall
Figure~\ref{fig_imp_tile_example}) or even local stiffnesses. Further,
local properties can also be prescribed at the tile level as ${\mathbf
    I}(u, v, w) : [0, 1]^3 \rightarrow \Reals^m$, where ${\mathbf I}^1$
defines the implicit geometry, with ${\mathbf I}^1(u, v, w) > 0$ being the
interior, and ${\mathbf I}^r(u, v, w)$, for coordinate $1 < r
\le m$, optionally prescribing a variety of local tile properties.

Properties that are described at the tile level are going to be
repeated throughout the lattice for all tiles.  If a tile's property is
periodic at the tile level\footnote{A property is considered
    \textit{periodic at the tile level} if it is identical for opposite tile
    boundaries; for example of same values for $u = 0$ and $u = 1$}, the
property will be periodic throughout the lattice.  In contrast,
properties that are prescribed by ${\mathcal T}$ will preserve the
continuity of the entire shape (following the continuity of ${\mathcal
    T}$) while being of possibly completely different values for different
tiles (locations).  See~\cite{Hong2023} for more on implicit tiles in
trivariate based lattices.

The design of lattices using implicit tiles entails several steps,
starting from the creation of implicit \Bspline{} trivariate tiles,
that is discussed in Section~\ref{subsec-create-implicit-tiles}. Then,
the synthesis of constant- and variably-controlled distance offsets of
implicits, in individual tile, is discussed in
Section~\ref{subsec-create-implicit-offset}. Mapping the tiles
through the \textit{macro-shape} ${\mathcal T}$, a constant offset will no
longer be preserved and in Section~\ref{subsec-prescribed-offsets}, we
also consider constant- and variably-controlled offsets in Euclidean
lattice space, after the mapping through \textit{macro-shape ${\mathcal T}$.}



\subsection{Creating implicit tiles}
\label{subsec-create-implicit-tiles}

We focus here on implicit trivariate \Bspline{} tiles that are created
in one of the following possible ways:
\begin{itemize}
    \item \textbf {Distance fields}.
    Consider a set of space curve and surface
    entities $E_q$ as $C_q(t_q)$ and $S_q(t_q, r_q)$, in $[0, 1]^3$.
    Let $d^{min}_q(p) = \min\limits_{t_q} ||p - E_q||$ be
    the minimum distance between space point $p$ and entity $E_q$.
    Then, let $D^{min}_q(p) = \min\limits_{q} d^{min}_q(p)$, or the
    minimum distance between $p$ and all entities
    $C_q(t_q)$ {\Gershon and} $S_q(t_q, r_q)$.

    Now consider a scalar \Bspline{} trivariate {\Gershon of general degrees
    $(d_u, d_v, d_w)$}:
    \begin{eqnarray}
        \lefteqn{\mbox{\hspace{-0.3in}} {\mathbf I}(u, v, w) = }\nonumber \\
        \mbox{\hspace{-0.3in}} &  &
        \sum_{i=0}^{n_u-1}
        \sum_{j=0}^{n_v-1}
        \sum_{k=0}^{n_w-1} p_{ijk}
        B_i^{d_u}(u) B_j^{d_v}(v) B_k^{d_w}(w)
        \label{eqn-trivar-bspline}
    \end{eqnarray}
    with $(n_u \times n_v \times n_w)$ coefficients and
    $u, v, w \in [0, 1]$.  Assuming a uniform spacing of coefficients
    in a 3D grid in $[0, 1]^3$, coefficient $p_{ijk}$ is located at
    $\left(\frac{i}{n_u-1}, \frac{j}{n_v-1}, \frac{k}{n_w-1}\right)$.

    Set $p_{ijk}$ to be $p_{ijk} =
    D^{min}_q\left(\frac{i}{n_u-1}, \frac{j}{n_v-1},
    \frac{k}{n_w-1}\right)$.  By spline approximation
    properties, ${\mathbf I}(u, v, w) = d_0$ is now an implicit
    \Bspline{} trivariate that approximates the iso-distance
    $d_0$ from $C_q(t_q)$ {\Gershon and} $S_q(t_q, r_q)$.

    Finally, $d^{min}_q(p) = \min\limits_{t_q} ||p - C_q(t_q)||$
    can be computed by seeking the minimum of its square, or its
    vanishing derivative parametric locations:
    \[
    0 = \frac{d (p - C_q(t_q))^2}{dt_q} = (p - C_q(t_q)) \cdot C_q'(t_q),
    \]
    solving a non-linear equation.
    $C^1$ discontinuities in $C_q(t_q)$ must also be examined for the
    minimum, as well as the two end points of $C_q$.
    The minimum point distance
    to surface $S_q(t_q, r_q)$ can be similarly derived.

    The tiles shown in Figure~\ref{fig_imp_tile_example} were created
    using this approach and are approximated and displayed using
    marching cubes, as an open tile (in
    Figure~\ref{fig_imp_tile_example}~(a)) and as closed tiles with
    boundary (Figure~\ref{fig_imp_tile_example}~(b) and ~(c)). Then,
    Figure~\ref{fig-tiles-dist-field} shows two additional examples
    that are also graded geometrically and material-wise (color).


    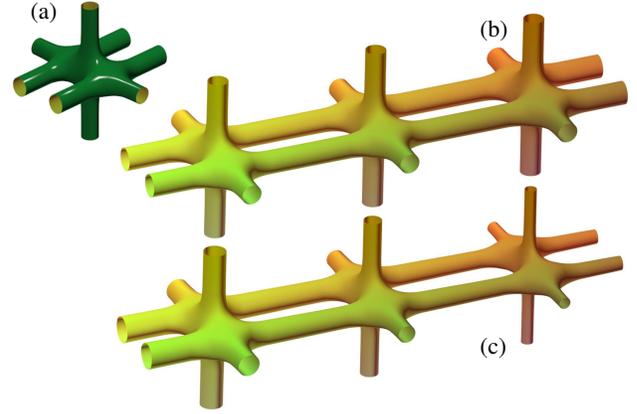
\begin{figure}
        \begin{center}
            \includegraphics[height=2.15in]{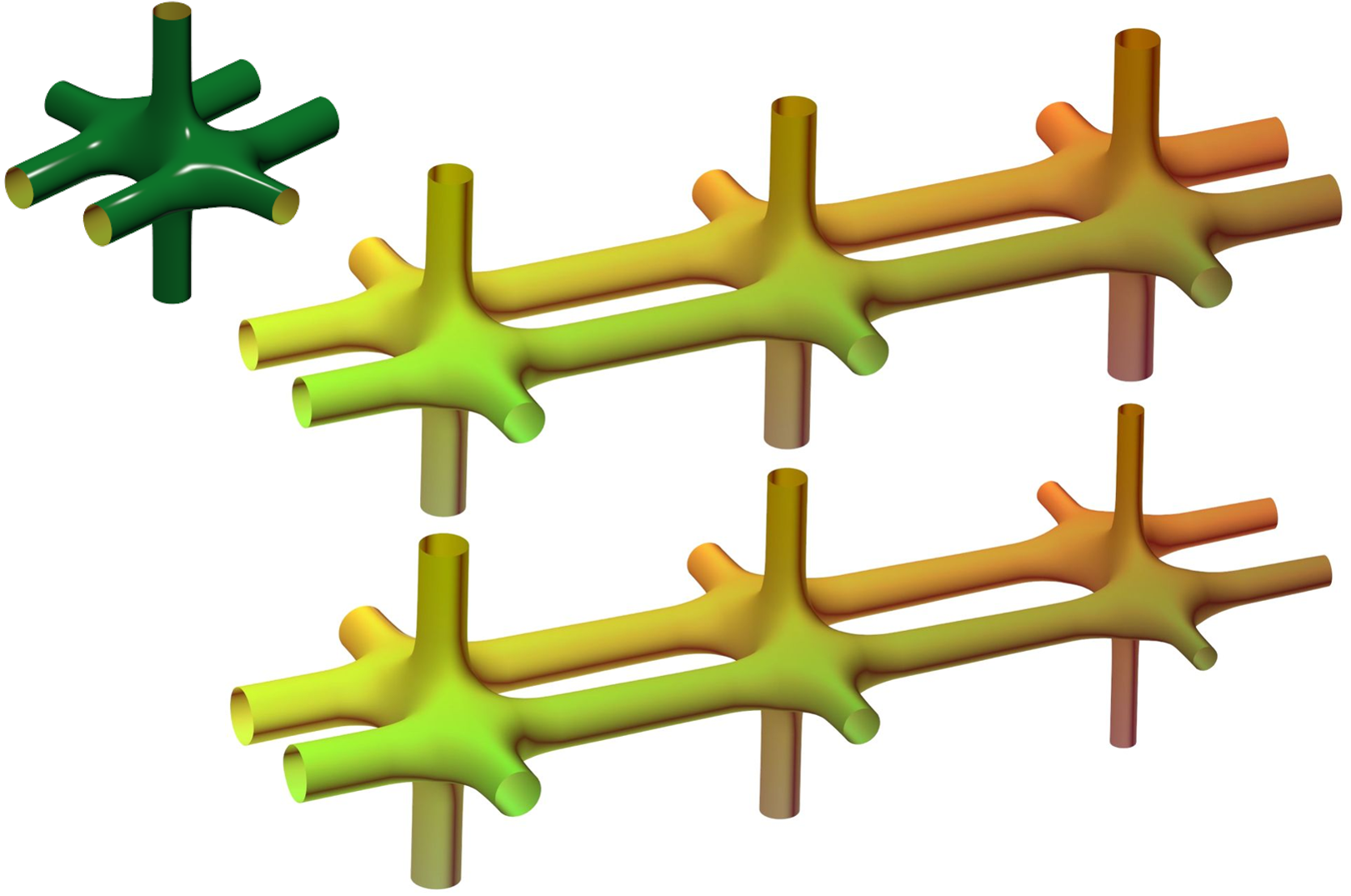} \\
            \begin{picture}(0,0)
                \put( -110, 160){(a)}
                \put(   60, 153){(b)}
                \put(   60,  33){(c)}
            \end{picture}
        \end{center}
        \mbox{\vspace{-0.55in}}\\[-0.55in]
        \caption{Two examples of three implicit tiles in a row.
		 The basic tile is shown in (a) built using the
		 distance field implicit tile creation approach, and
		 emply four orthogonal lines (two of which are, in
		 fact, parallel).
            	 The tiles are created uniform geometrically in (b)
            	 and graded geometrically in (c).  Both are created
            	 heterogeneous with respect to material (color).  Each
            	 tile is a tri-quadratic \Bspline{} with 10 control
            	 points in each direction.}
        \label{fig-tiles-dist-field}
        \mbox{\vspace{-0.5in}}\\[-0.5in]
    \end{figure}

    \item \textbf{Implicit Evaluation}.  Let ${\mathcal I}(u, v, w) = 0$ be
    an implicit function, for $u, v, w \in [0, 1]$.  Reconsider
    ${\mathbf I}(u, v, w)$ from
    Equation~\eqnref{eqn-trivar-bspline}, and set $p_{ijk}$ to
    equal $p_{ijk} = {\mathcal I}\left(\frac{i}{n_u-1},
    \frac{j}{n_v-1}, \frac{k}{n_w-1}\right)$.  By the properties
    of Bernstein/\Bspline{} basic functions, ${\mathbf I}(u, v, w)$
    approximates ${\mathcal I}$ (uniformly, as $(n_u, n_v, n_w)$
    converge to infinity).

    In other words, given any implicit function ${\mathcal I}(u, v,
    w) = 0$, by simply sampling ${\mathcal I}$, one can approximate
    ${\mathcal I}$ as closely as desired, via an increase in the
    mesh size, as ${\mathbf I}(u, v, w) = 0$.
    Figure~\ref{fig-tile-eval-gyroid} shows an
    example of sampling a TPMS Gyroid function into an implicit
    tri-quadratic \Bspline{} function of size $10 \times 10
    \times 10$.

    \begin{figure}
        \begin{center}
            \includegraphics[height=1.75in]{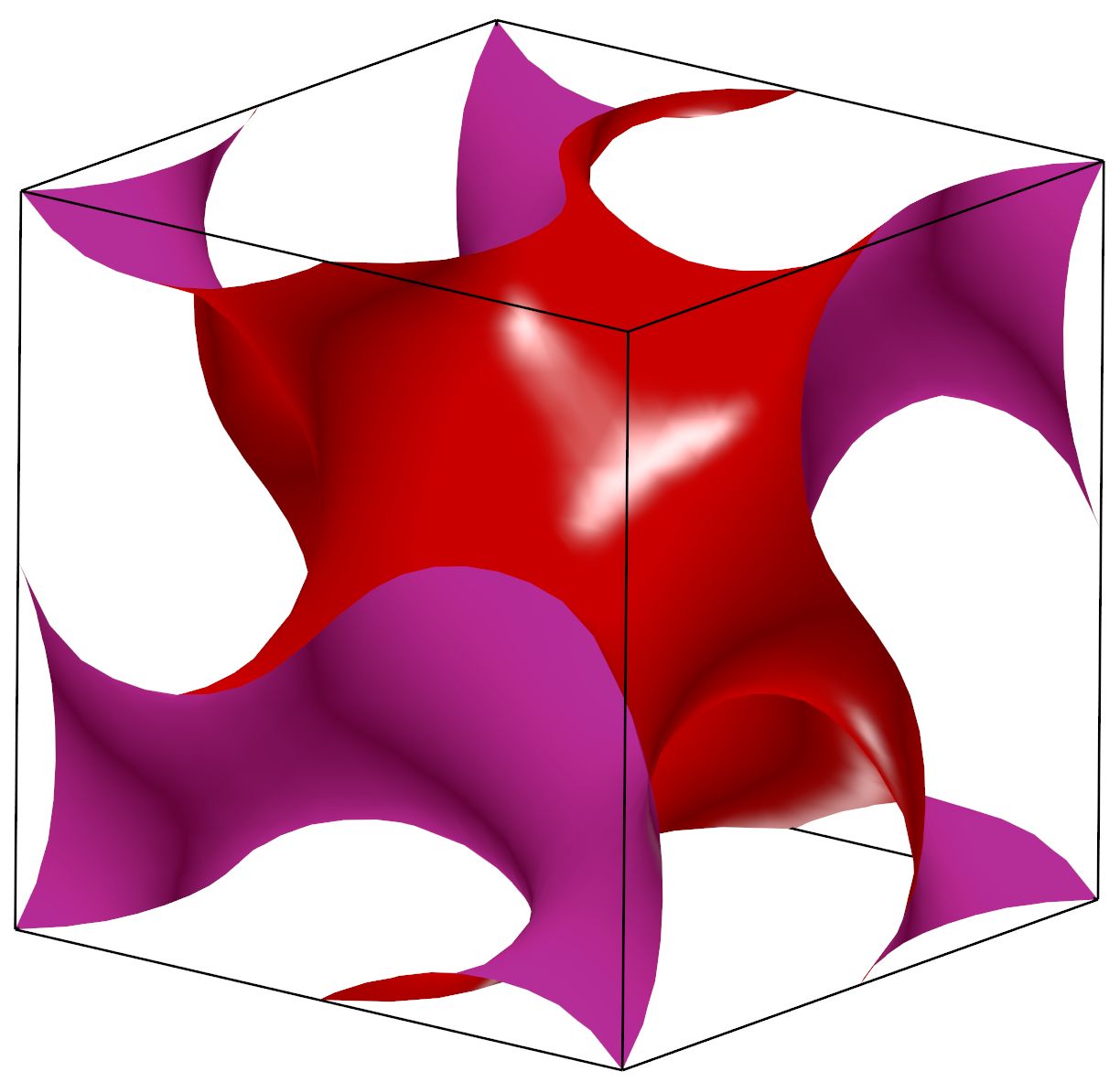}
        \end{center}
        \mbox{\vspace{-0.4in}}\\[-0.4in]
        \caption{An implicit tri-quadratic \Bspline{} function of
            size $10 \times 10 \times 10$ approximating a
            Gyroid function tile by sampling.}
        \label{fig-tile-eval-gyroid}
    \mbox{\vspace{-0.6in}}\\[-0.6in]
    \end{figure}

    \item \textbf{Implicit least squares}.  As an alternative to Implicit
    Evaluation, one can aim to best fit the \Bspline{} trivariate
    to a given implicit function ${\mathcal I}(u, v, w) = 0$.

    In this method, the coefficients $p_{ijk}$'s in ${\mathbf
        I}(u,v,w)$ from Equation~\eqnref{eqn-trivar-bspline} are
    determined by minimizing $E$, the sum of squared distances
    between ${\mathbf I}(u,v,w)$ and ${\mathcal I}(u,v,w)$, formulated
    as follows.  \[ E = \sum_{m=1}^{M} ({\mathbf I}(u_m, v_m, w_m) -
    {\mathcal I}(u_m, v_m, w_m))^2 \] where $M$, being larger than
    the number of coefficients, is the number of point samples
    for fitting, and $u_m$, $v_m$, $w_m$ are in $[0,1]$.  The
    least square problem is formulated by finding $p_{ijk}$'s
    that satisfy $\frac{\partial E}{\partial {\mathbf p}_{ijk}} = 0$
    for all $i$, $j$, and $k$, or
    \begin{eqnarray}
        \sum_{m=1}^{M}({\mathbf I}(u_m, v_m, w_m) -
          {\mathcal I}(u_m, v_m, w_m)) \cdot
          \frac{\partial {\mathbf I}(u_m, v_m, w_m)}
                              {\partial {\mathbf p}_{ijk}} = 0,\nonumber
    \end{eqnarray}
    and can be solved via direct methods.

    Hu and Lin~\cite{Hu2021} proposed the Least Square
    Progressive and Iterative Approximation (LSPIA) method to
    fit a scalar \Bspline{} trivariate to the implicit function in
    an iterative manner.  This method initializes $p_{ijk}$'s to
    zero and updates the values of $p_{ijk}$'s iteratively until
    they converge, by distributing the error of the $m$-th sample to
    the coefficients in ${\mathbf I}(u,v,w)$ that contribute to
    evaluation of the sample, with a weight of
    $B_i^{d_u}(u_m)B_j^{d_v}(v_m)B_k^{d_w}(w_m)$.

\end{itemize}

In all these tiles' construction schemes, continuity must be preserved
between adjacent trivariate tiles.  Hence, typically, for $C^0$
continuity, open-end conditions can be employed and the coefficients
of adjacent trivariates are made identical on the shared boundary
faces.  Further, if the offset is constant, a simple alternative is to
impose periodic boundary conditions on the fitted spline.


\subsection{The Creation of variable distance offsets of Implicits}
\label{subsec-create-implicit-offset}

Consider the implicit tile ${\mathbf I}(u, v, w)$ and let the steepest
ascent/descent direction of ${\mathbf I}$, be $(V_u, V_v, V_w) =
\frac{\nabla {\mathbf I}}{|\nabla {\mathbf I}|}$, where ${\nabla {\mathbf I}}$ is
the gradient of ${\mathbf I}$. Then, using a first order Taylor
series expansion,

\begin{eqnarray}
    \lefteqn{{\mathbf I}(u+dV_u, v+dV_v, w+dV_w)} \nonumber\\
    & = & {\mathbf I}(u,v,w) + \frac{\partial {\mathbf I}}{\partial u}dV_u +
    \frac{\partial {\mathbf I}}{\partial v}dV_v +
    \frac{\partial {\mathbf I}}{\partial w}dV_w +
    O(d^2) \nonumber\\
    & \simeq & {\mathbf I}(u,v,w) + d \left( \frac{\partial {\mathbf I}}{\partial u},
    \frac{\partial {\mathbf I}}{\partial v},
    \frac{\partial {\mathbf I}}{\partial w} \right) \cdot
    (V_u, V_v, V_w) \nonumber\\
    & = & {\mathbf I}(u,v,w) + d \nabla {\mathbf I} \cdot
    \frac{\nabla {\mathbf I}}{|\nabla {\mathbf I}|}.
    \label{eqn-imp-offset}
\end{eqnarray}

Now $d$ can prescribe (a first order approximation of) the offset
distance.  Further, $d$ can be a function of $u, v, w$ as $d(u, v, w)$
and hence Equation~\eqnref{eqn-imp-offset} can also represent variable
distance inter-tiles offset but also intra-tile offset!  Then, ${\mathbf
    I}_d$, a first order approximation of the variable distance offset of
${\mathbf I}$ by $d(u, v, w)$ can now be represented as the new implicit

\begin{equation}
    {\mathbf I}_d(u, v, w)
    = {\mathbf I}(u, v, w) + d(u, v, w) \frac{\nabla {\mathbf I} \cdot \nabla {\mathbf I}}
    {|\nabla {\mathbf I}|}.
    \label{eqn-imp-offset2}
\end{equation}

Now ${\mathbf I}(u, v, w)$ and ${\mathbf I}_d(u, v, w)$ form together a new
tile of a certain, possibly variable, thickness.  Considering the
material between ${\mathbf I}(u, v, w)$ and ${\mathbf I}_d(u, v, w)$ and
we have formed a tile of (variable) controlled wall thickness.
Further and alternatively, one can employ ${\mathbf I}_{-d}(u, v, w)$ and
${\mathbf I}_{+d}(u, v, w)$ to achieve a symmetric tile around
${\mathbf I}(u, v, w)$.

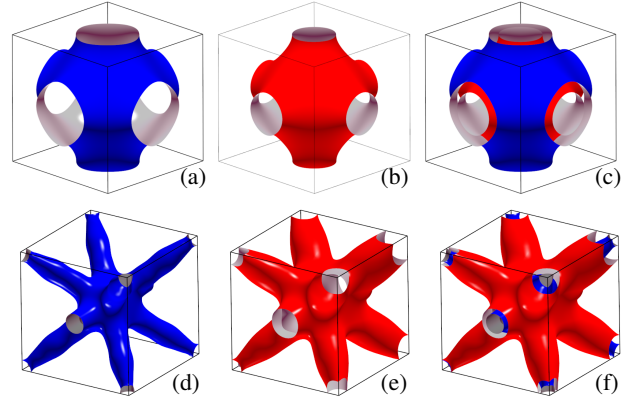
\begin{figure}
    \includegraphics[height=1.05in]{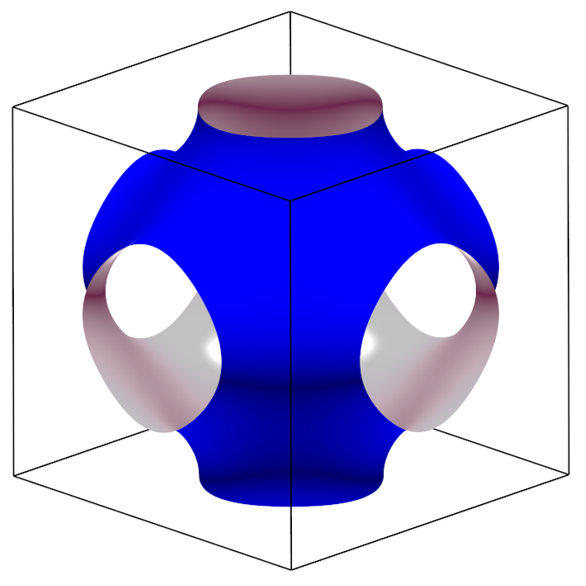}
    \includegraphics[height=1.05in]{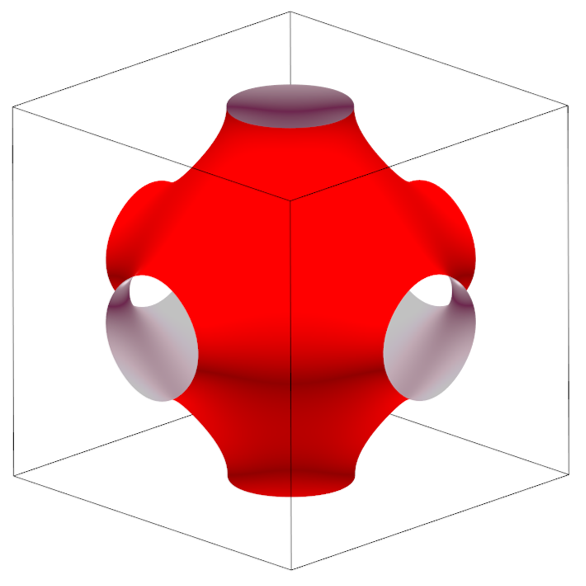}
    \includegraphics[height=1.05in]{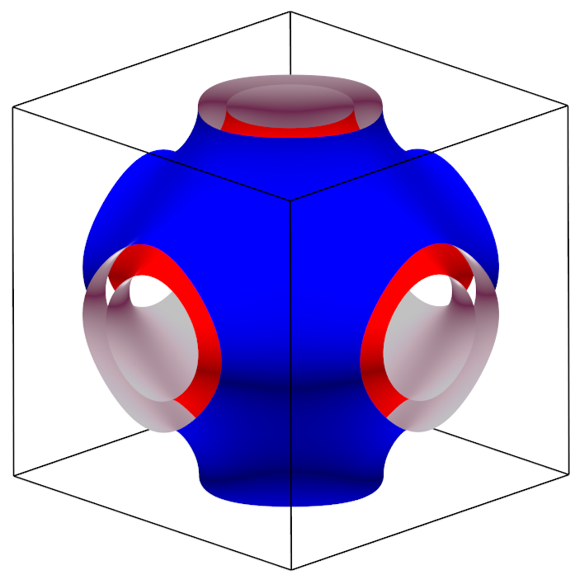}\\
    \includegraphics[height=1.07in]{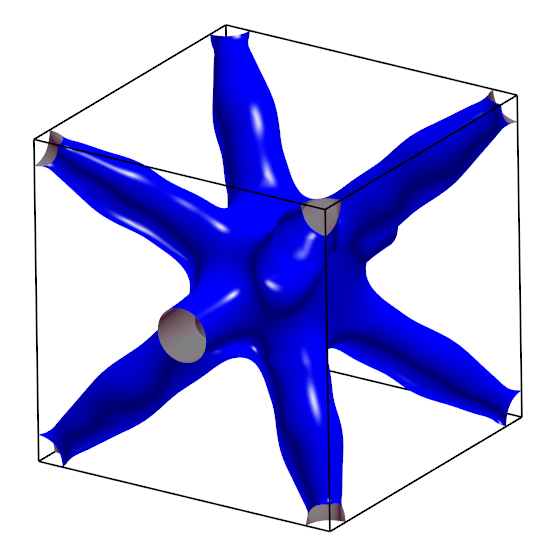}
    \includegraphics[height=1.07in]{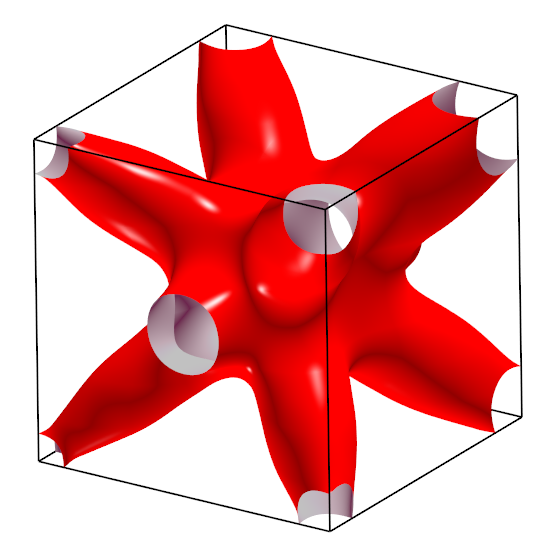}
    \includegraphics[height=1.07in]{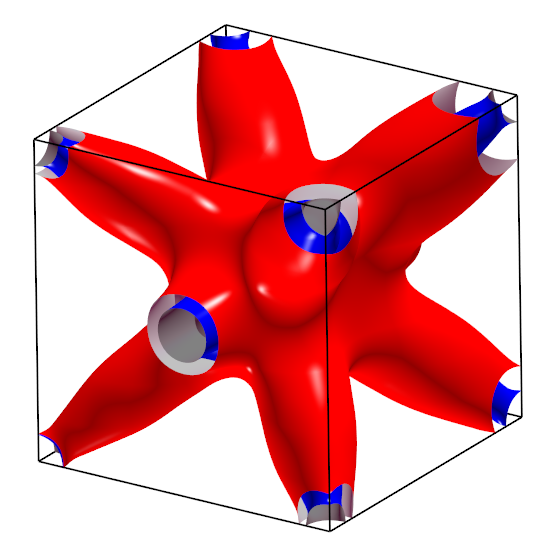} \\
    \begin{picture}(0,0)
        \put(  65, 95){(a)}
        \put( 141, 95){(b)}
        \put( 222, 95){(c)}
        \put(  62, 17){(d)}
        \put( 141, 17){(e)}
        \put( 222, 17){(f)}
    \end{picture}
    \mbox{\vspace{-0.4in}}\\[-0.4in]
    \caption{An implicit {\Gershon trivariate TPMS (Schwarz-P)
	approximation tile (created using an interpolatory fitting scheme
	similar to~\cite{Hu2021})} in (a) and an implicit {\Gershon trivariate}
        3D cross tile {\Gershon (created using a distance field to curves)}
	 in (d) are employed to demonstrate
        (the zero sets of the) variable distance offset
        computations, as are shown in (b) and (e).
        (c) and (f) show (the zero sets of the) implicits of the
        base as well as the variable distance offset implicits,
        in the same images.}
    \label{fig-tile-simple-example-offset}
    \mbox{\vspace{-0.5in}}\\[-0.5in]
\end{figure}

Figure~\ref{fig-tile-simple-example-offset} shows two examples, of an
implicit TPMS (Schwarz-P) tile in
Figure~\ref{fig-tile-simple-example-offset}~(a) and of an implicit
\Bspline{} trivariate in (d).  (b) and (e) show the results of
(the zero sets of the) variable distance offsets computed using
Equation~\eqnref{eqn-imp-offset2}, whereas (c) and (f) show
together {\MyungSoo (the zero sets of) the} base implicit tiles and their
respective variable distance offset implicits, in the same images.

\subsection{Constant-/variably-controlled offsets in lattice space}
\label{subsec-prescribed-offsets}

While in the previous section, we presented an approach to
compute constant- and variably-controlled offsets of implicits, by
mapping the offset implicit through the \textit{macro-shape} ${\mathcal T}$,
the offset distances are not likely to be preserved.  Further, by
computing offsets of individual tiles that are $C^0$ continuous,
offsets of those tiles will no longer be continuous.  Herein, we aim
to alleviate these difficulties by:
\begin{itemize}
    \item Estimations of offset distances in Euclidean space and after the
    mapping through ${\mathcal T}$ took place.
    \item Fitting a single implicit spline for the entire lattice,
    ensuring any desired continuity throughout.
    \item Topological changes in the tiles and throughout the lattice could be
    accommodated in a simple way, including due to infinitesimal
    offset changes.
\end{itemize}


\begin{figure*}
    \begin{center}
        \includegraphics[height=4.1in]{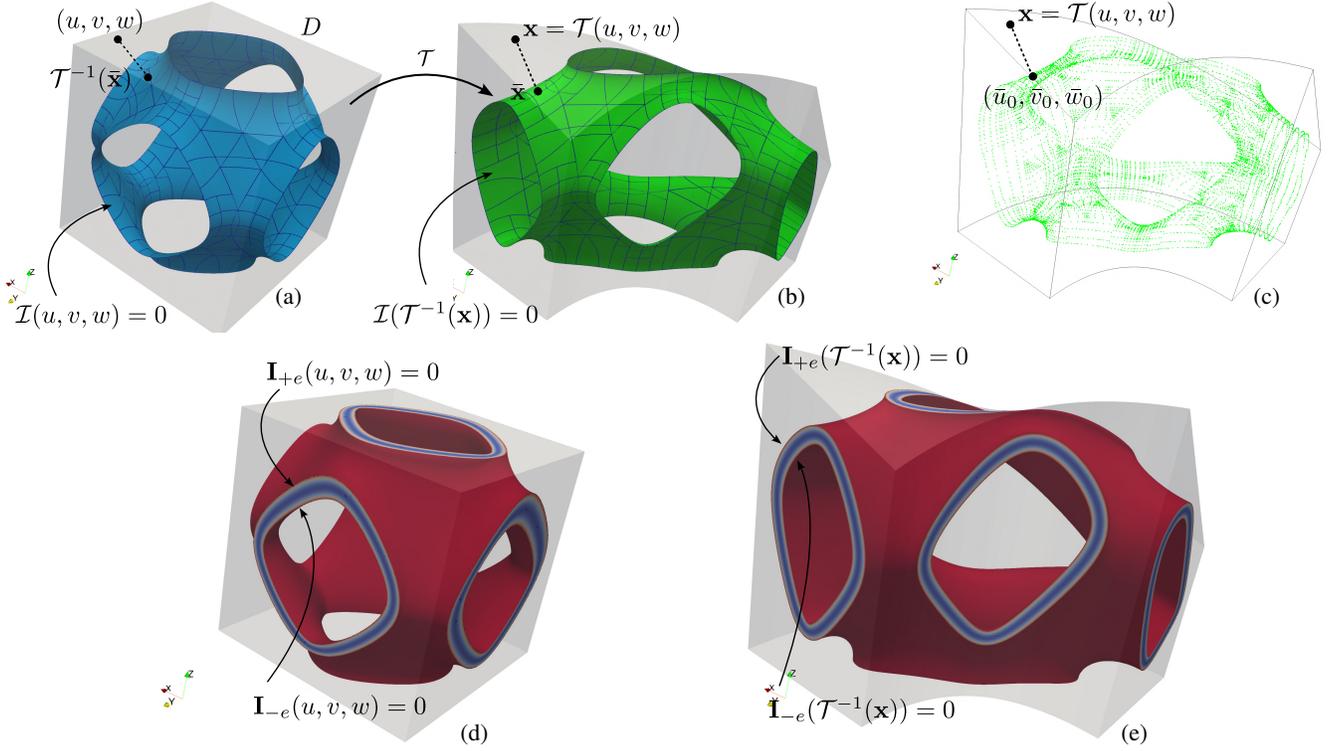}
        \begin{picture}(0,0)
            \put( -140, 180){(a)}
            \put(   50, 180){(b)}
            \put(  230, 180){(c)}
            \put(  -70,  15){(d)}
            \put(  180,  15){(e)}
        \end{picture}
    \end{center}
    \mbox{\vspace{-0.4in}}\\[-0.4in]
    \caption{\Pablo Creation of controlled offsets for a Schwarz-P tile in lattice space.
      In (a), the tile $\mathcal{I}(u,v,w)=0$ in $D$ is represented (through a non-conforming high-order reparatemerization).
      In (b), the tile is mapped by a trivariate $\mathcal{T}$.
      A cloud of points lying on the mapped tile is represented in (c): these points are used for accelerating
      the search of the closest point $\bar{\mathbf{x}}$ to a given point $\mathbf{x}$.
      Finally, the material between $\mathbf{I}_{-e}$ and $\mathbf{I}_{+e}$ is represented
      in the domain $D$ (in (d)) and in the lattice space in (in (e)).}
    \label{euclidean-lattice}
    \mbox{\vspace{-0.35in}}\\[-0.35in]
\end{figure*}

For some specific tile ${\mathcal I}(u,v,w)$ in ${\mathcal T}$, ${\mathcal T}(t_x,
t_x, t_z)$ is ${\mathcal T}(u, v, w)$ as $(u,v,w)$ and $(t_x, t_x, t_z)$
equate.  Thus, given a generic point ${\mathbf x} = {\mathcal T}(t_x, t_x,
t_z) = {\mathcal T}(u, v, w)$, we denote as $\bar{\mathbf x}$ its closest
point such that its pre-image {\MyungSoo lays} on the surface ${\mathcal I}(u,v,w) =
0$, i.e., ${\mathcal I}({\mathcal T}^{-1}(\bar{\mathbf x})) = 0$ {\Pablo (see Figures~\ref{euclidean-lattice} (a)-(b))}.
The computation
of $\bar{\mathbf x}$ will be addressed below.  In this way, the new
geometry is defined as region between the implicit functions ${\mathbf
    I}_{+e}$ and ${\mathbf I}_{-e}$, defined as:
\begin{subequations}
    \begin{align}
        {\mathbf I}_{+e}(u, v, w) &= \signdist(u,v,w) + d\,,\\
        {\mathbf I}_{-e}(u, v, w) &= \signdist(u,v,w) - d\,,
    \end{align}
\end{subequations}
where $\signdist(u,v,w)$ is the signed distance between the points ${\mathbf x}$ and $\bar{\mathbf x}$, and it is defined as
\begin{equation}
    \signdist(u,v,w) = \text{sign}({\mathcal I}(u,v,w))||{\mathcal T}(u, v, w) - \bar{\mathbf x}(u,v,w)||.
\end{equation}
{\Pablo The resulting functions ${\mathbf I}_{+e}$ and ${\mathbf I}_{-e}$ for a Schwarz-P tile example
is shown are Figure~\ref{euclidean-lattice} (d), and after composition with a trivariate function $\mathcal{T}$
in Figure~\ref{euclidean-lattice} (e).}
As before, the offset distance $d$ with respect to the mid-surface
${\mathcal I}(u,v,w) = 0$ can be a function of $u,v,w$ and can be chosen
separately for ${\mathbf I}_{+e}$ and ${\mathbf I}_{-e}$.
{\Pablo Actually, it can be appreciated that to achieve a constant offset in Figure~\ref{euclidean-lattice} (e),
the offsets in the domain $D$ (Figure~\ref{euclidean-lattice} (d)) are non constant.}
The functions ${\mathbf
    I}_{+e}$ and ${\mathbf I}_{-e}$ can be further approximated as \Bspline{}
trivariates by using any of the methods proposed in
Section~\ref{subsec-create-implicit-tiles}.

The question of finding the closest point $\bar{\mathbf x}$ remains open.
For a given point ${\mathbf x}(u,v,w)$, this can be formulated as a
constrained minimization problem:
    \begin{equation}
        \begin{split}
            (\bar u, &\bar v,\bar w) = \underset{t_x,t_y,t_z}{\text{arg min}}\; || {\mathcal T}(t_x,t_y,t_z) - {\mathbf x}||\\
            &\text{ s.t. }{\mathcal I}(\bar u,\bar v,\bar w) = 0 \text{ and } \bar{\mathbf x} = {\mathcal T}(\bar u,\bar v,\bar w)\,.
        \end{split}
    \end{equation}
For solving such constrained optimization problem, similarly to \cite{saye2014high}, we recast it as: Find $(\bar u,\bar v,\bar w, \bar \mu)$ such that
\begin{equation}
    (\bar u,\bar v,\bar w, \bar \mu) = \underset{t_x,t_y,t_z,\lambda}{\text{arg min}}\; f(t_x,t_y,t_z,\lambda)\,,
\end{equation}
where $f:\Reals^3\times\Reals\to\Reals$ is the functional given by\footnote{The method originally proposed in~\cite{saye2014high} can be seen as a particularization for the case in which the macro-shape ${\mathcal T}$ is the identity.}:
\begin{equation}
    f(t_x,t_y,t_z, \lambda) = \frac{1}{2}||{\mathcal T}(t_x,t_y,t_z) -  {\mathbf x}||^2 + \lambda {\mathcal I}(t_x,t_y,t_z)\,,
\end{equation}
and $\lambda$ is a Lagrange multiplier that constrains the closest point to be in the zero level set of $\mathcal{I}(t_x,t_y,t_z)$.
The saddle-point of this functional is sought by means of an iterative Newton-Raphson algorithm starting from an initial guess ${\mathbf w}_0=(\bar u_0,\bar v_0,\bar w_0, \bar \mu_0)$, where successive approximations are computed as:
\begin{equation}
    {\mathbf w}_{i+1} = {\mathbf w}_{i} - \left(\nabla^2 f({\mathbf w}_{i})\right)^{-1} \nabla f({\mathbf w}_{i})\,.
\end{equation}
The functional's gradient is:
\begin{equation}
    \nabla f =
    \begin{pmatrix}
        \left({\mathcal T}(t_x,t_y,t_z) -  {\mathbf x}\right) \nabla {\mathcal T}(t_x,t_y,t_z) + \lambda \nabla {\mathcal I}(t_x,t_y,t_z) \\
        {\mathcal I}(t_x,t_y,t_z)
    \end{pmatrix}
\end{equation}
and its Hessian:
\begin{equation}
    \nabla^2 f =
    \begin{pmatrix}
        \nabla^2 f_{t_x,t_y,t_z} & \nabla {\mathcal I}(t_x,t_y,t_z) \\
        \nabla {\mathcal I}(t_x,t_y,t_z) & 0
    \end{pmatrix}\,,
\end{equation}
where
\begin{equation}
    \nabla^2 f_{x,y,z} = \nabla^\top {\mathcal T}(t_x,t_y,t_z) \nabla {\mathcal T}(t_x,t_y,t_z) - {\mathbf x} \nabla^2 {\mathcal T}(t_x,t_y,t_z)\,.
\end{equation}

The choice of the initial guess ${\mathbf w}_0$ is critical for the
success and performance of the iterative process.  For that purpose,
we compute $(\bar u_0,\bar v_0,\bar w_0)$ by finding the closest point
to ${\mathbf x}$ in a cloud of $N$ precomputed points ${\mathbf
Y}=\left\lbrace {\mathbf y}_q\in\Reals^3 \;\vert\; {\mathcal
I}\left({\mathcal T}^{-1}({\mathbf y}_q) \right) = 0
\right\rbrace_{q=1}^{N}$ {\Pablo (see Figure~\ref{euclidean-lattice} (c))}.
The points ${\mathbf y}_q$ are created by
sampling the zero level-set of ${\mathcal I}$ in the domain $D$ and
then mapping them through ${\mathcal T}$.  The search of the closest
point in the cloud ${\mathbf Y}$ is performed by means of the k-d tree
algorithm proposed in~\cite{saye2014high}.  The sampling of the zero
level-set of ${\mathcal I}$ in $D$ is performed by means of the
algorithms proposed in~\cite{saye2022high,saye2015high} for the
generation of quadrature rules for implicit domains defined through
polynomials and general $C^1$ analytic functions, respectively.  Both
algorithms, as well as the k-d tree mentioned above, are implemented
in the open source project \texttt{algoim}~\cite{algoim}.

It is worth mentioning that, when placing several implicit tiles
together in the domain $D$ (recall Figure~\ref{fig-tiles-dist-field}),
for a given point $\mathbf x$ belonging to the subdomain of one of the
tiles, the closest point $\bar{\mathbf x}$ may not belong to the same
subdomain, but to the one of a neighboring tile.  For this reason, in
the procedure detailed above, the implicit function
$\mathcal{I}(x,y,z)$ must be understood as the union of the functions
of all the tiles considered, that are assumed to be smooth enough at
their interfaces.  In addition, when approximating the resulting
functions ${\mathbf I}_{+e}$ or ${\mathbf I}_{-e}$ as \Bspline{}
trivariates, the whole set of tiles is considered together, fitting
one single \Bspline{} for the whole domain. The order, knot sequences,
and number of control points of such spline approximations can be
chosen to guarantee the required continuity at the tiles' interfaces.


\section{Direct Simulation of Implicit lattices}
\label{sec-analysis-implicit}

While the geometries presented above are analysis suitable,
the application of traditional boundary fitted finite element
methods~\cite{Hughes2000} for solving Partial Differential Equations
(PDEs) may be cumbersome as it requires the creation of high-quality
boundary-fitted parametric meshes that may be a non-trivial task.  In
order to overcome this difficulty, in this work we leverage on
unfitted discretization
techniques~\cite{cutfem,duester2008,main2018,badia2018}.

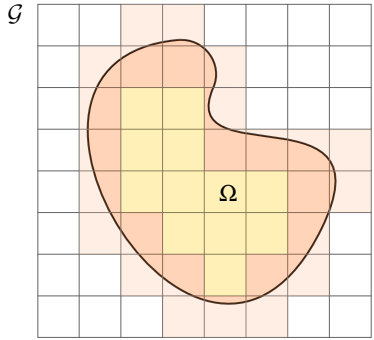
\begin{figure}
    \begin{center}
        \includegraphics[height=1.75in]{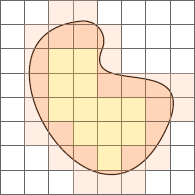}
        \begin{picture}(0,0)
            \put( -60, 52){$\Omega$}
            \put( -140, 120){$\mathcal{G}$}
        \end{picture}
    \end{center}
    \mbox{\vspace{-0.30in}}\\[-0.30in]
    \caption{In unfitted finite elements methods, the computational domain to analyze $\Omega$ is immersed in a background grid $\mathcal{G}$, independent from the geometry, that is used to discretize the PDE solution. The elements of the grid can be classified as internal (yellow), cut (orange), or external (white).}
    \label{fig-cutfem}
    \mbox{\vspace{-0.5in}}\\[-0.5in]
\end{figure}

The main idea of such family of methods is to immerse the domain
$\Omega$ in a background mesh $\mathcal{G}$ (see
Figure~\ref{fig-cutfem}) that serves as base for discretizing the PDE
at hand, instead of creating a boundary-fitted discretization.
Thus, following a Galerkin setting, the degrees-of-freedom and basis functions of the PDE discretized solution and test functions are built upon the background mesh $\mathcal{G}$.
In our case, $\mathcal{G}$ is chosen as (a refinement of) the parametric domain partition of the implicit \Bspline{} trivariates.
Thus, $\mathcal{G}$ is naturally split into three disjoint sets of elements, namely:
\begin{subequations}
    \begin{align}
        \mathcal{G}_{\text{in}} &= \left\lbrace Q \;\vert\, Q\in\mathcal{G}\;:\;Q\cap\Omega=Q\right\rbrace,\\
        \mathcal{G}_{\text{out}} &= \left\lbrace Q \;\vert\, Q\in\mathcal{G}\;:\;Q\cap\Omega=\emptyset\right\rbrace,\\
        \mathcal{G}_{\text{cut}} &= \left\lbrace Q \;\vert\, Q\in\mathcal{G}\;:\;Q\not\in\mathcal{G}_{\text{in}}\cup\mathcal{G}_{\text{out}} \right\rbrace.
    \end{align}
\end{subequations}
These three families correspond to the internal, external, and cut elements, respectively (recall Figure~\ref{fig-cutfem}), being the PDE at hand discretized by means of finite-element techniques solely in the active part of the grid, i.e, in the internal and cut elements.
Further details about the formulation of the method and the problems associated with unfitted techniques, including the treatment of cut elements, are deferred to {\MyungSoo Appendix}~\ref{sec-appendix}.

\section{Experimental Results}
\label{sec-results}

We now present some examples of creating lattices of implicit tiles with
constant and graded wall thickness as well as analyzing them.

\subsection{Prescribed offsets in the domain of the macro-shape}

\begin{figure*}
    \begin{center}
        \includegraphics[height=2.5in]{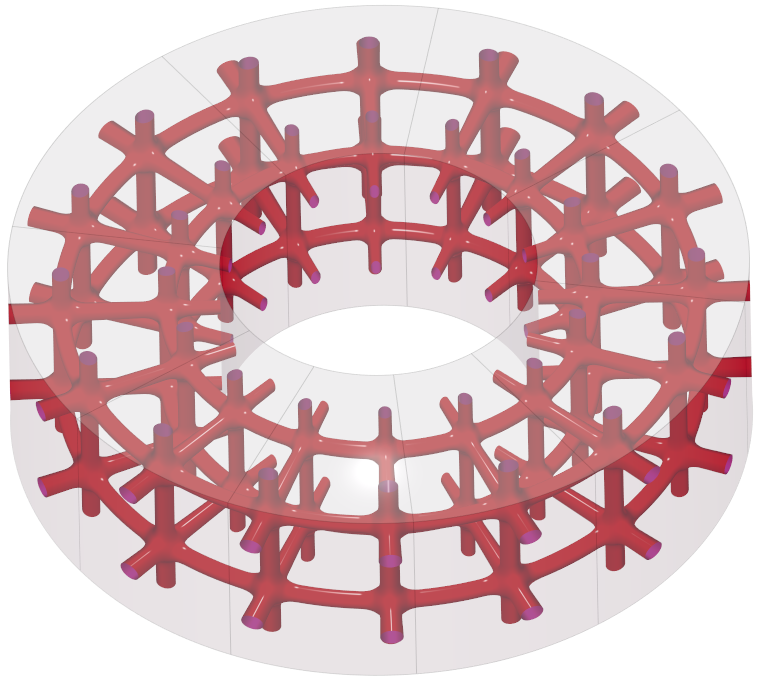}
        \includegraphics[height=2.5in]{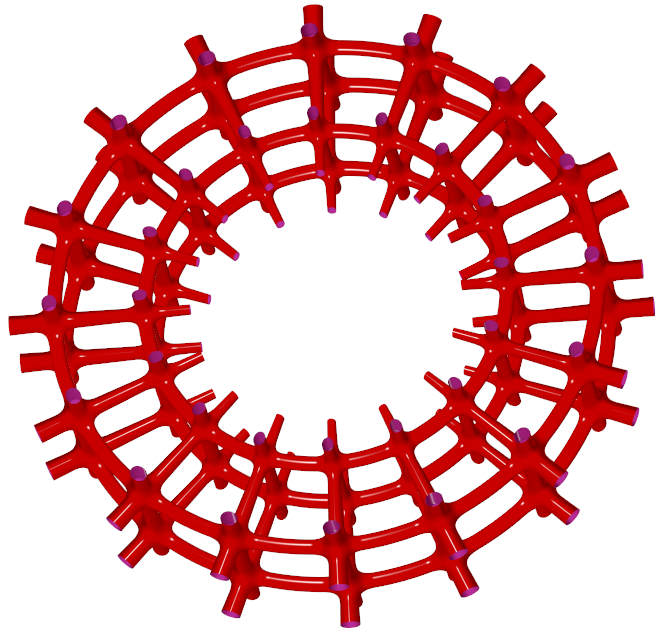} \\
        \includegraphics[height=2.5in]{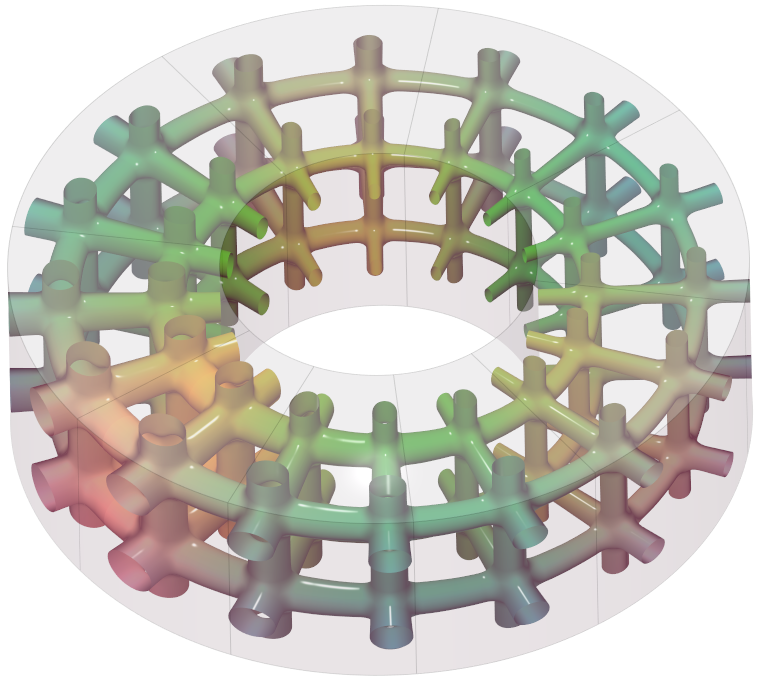}
        \includegraphics[height=2.5in]{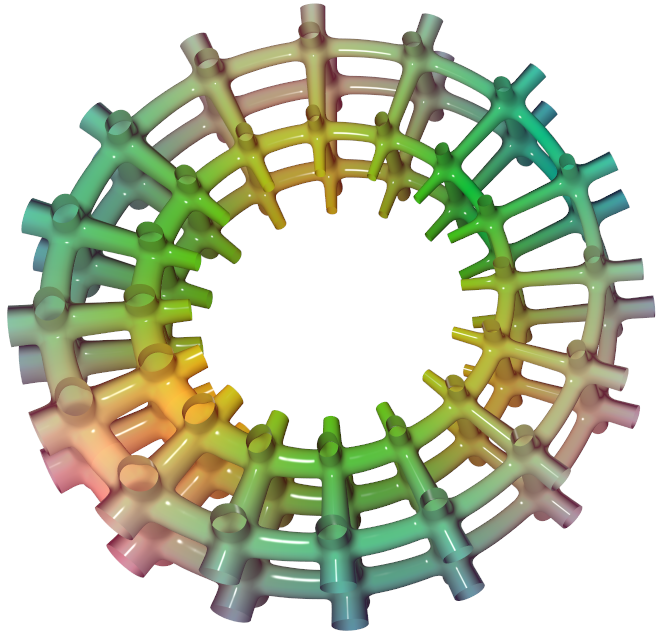}
        \begin{picture}(0,0)
            \put( -430, 190){(a)}
            \put(  -25, 190){(b)}
            \put( -430,  12){(c)}
            \put(  -25,  15){(d)}
        \end{picture}
    \end{center}
    \mbox{\vspace{-0.4in}}\\[-0.4in]
    \caption{An embedding of an implicit \Bspline{} 3D cross tile in a
        parametric \Bspline{} torus with a square cross section, $2
        \times 2 \times 16$ times (torus is shown translucent in (a)
        and (c)).  In (a) and (b) a uniform thickness and uniform
        material (color) lattice is presented where as in (c) and
        (d), the lattice varies in both the geometry and material,
        \QYoun{independently.}}
    \label{fig-tiles-cross-imp-diags-axes}
    \mbox{\vspace{-0.35in}}\\[-0.35in]
\end{figure*}
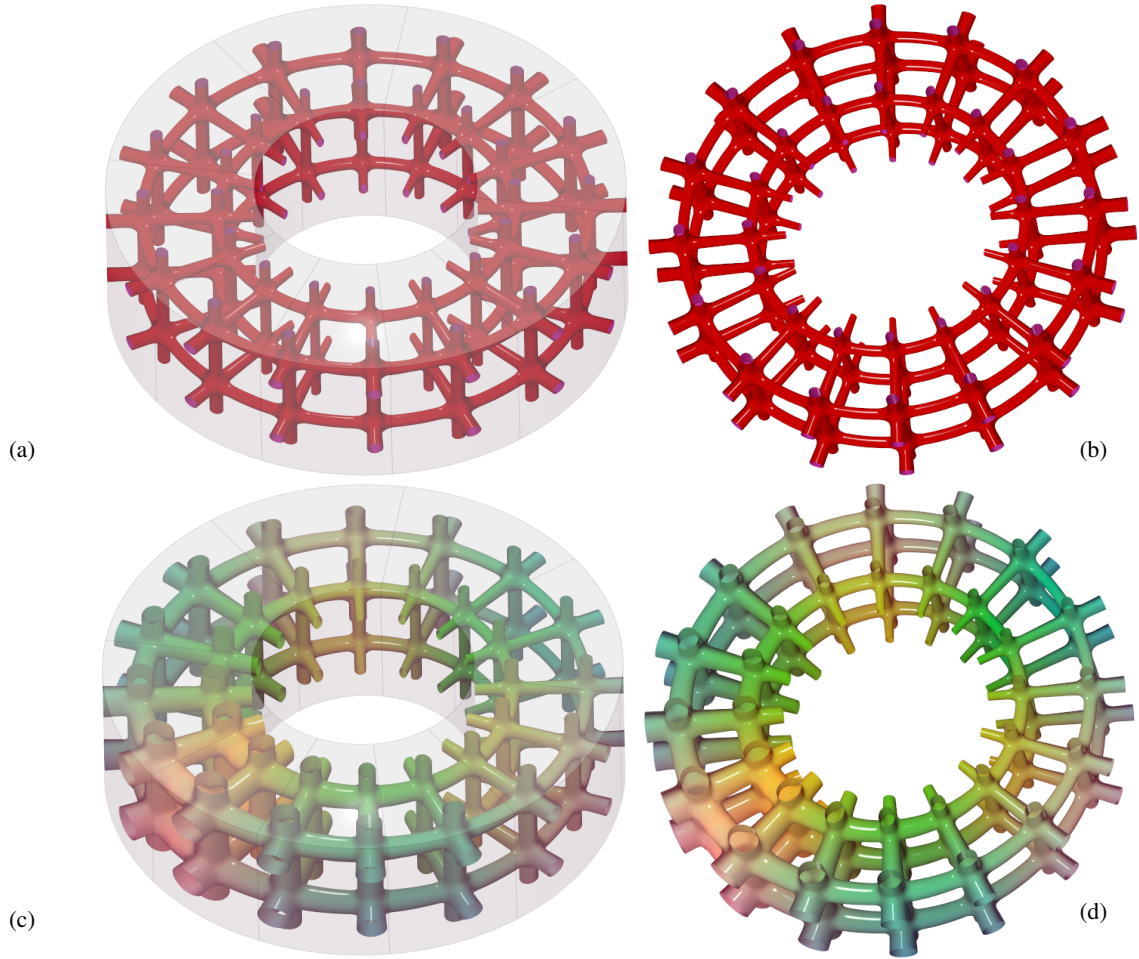

This sub-section portrays results that employ the offset computation
scheme presented in Section~\ref{subsec-create-implicit-offset}.
Figure~\ref{fig-tiles-cross-imp-diags-axes} presents a
constant as well as variable size offsets of a cross tile in a torus
macro-shape with a square cross section.  In (a) and (b) constant
offset tiles are shown, that after mapping through the macro-shape
${\mathcal T}$ the tiles are made thicker toward the outside. (c) and (d)
present a case where the thickness of the cross tiles are further
varying around the torus.  Note (c) and (d) also present graded
material content (colors) around the torus.  The cross tile was an
implicit trivariate of orders $(3\times3\times3)$ and
$(10\times10\times10)$ control points, and the macro shape is of
orders $(4\times2\times2)$ and $(13\times2\times2)$ control points in
$R^7$ where the first three dimensions prescribes the geometry,
three for the RGB color properties of the macro shape ${\mathcal T}$,
and the last dimension prescribes the offset distance $d$ from
Equation~\eqnref{eqn-imp-offset}.  The computation times of the
marching cube procedure took little over 10 seconds and little over 20
seconds for constant and variable distance offset tiles, respectively,
on an Intel Core i7-12700 2.10GHz PC with 12 threads. It took longer
to compute variable distance offsets than constant offsets because the
marching cubes are executed as many times as the number of tiles in
variable distance offsets, whereas the marching cubes' result can be
shared between the tiles, for constant offset tiles.

\begin{figure*}
    \begin{center}
        \includegraphics[height=1.7in]{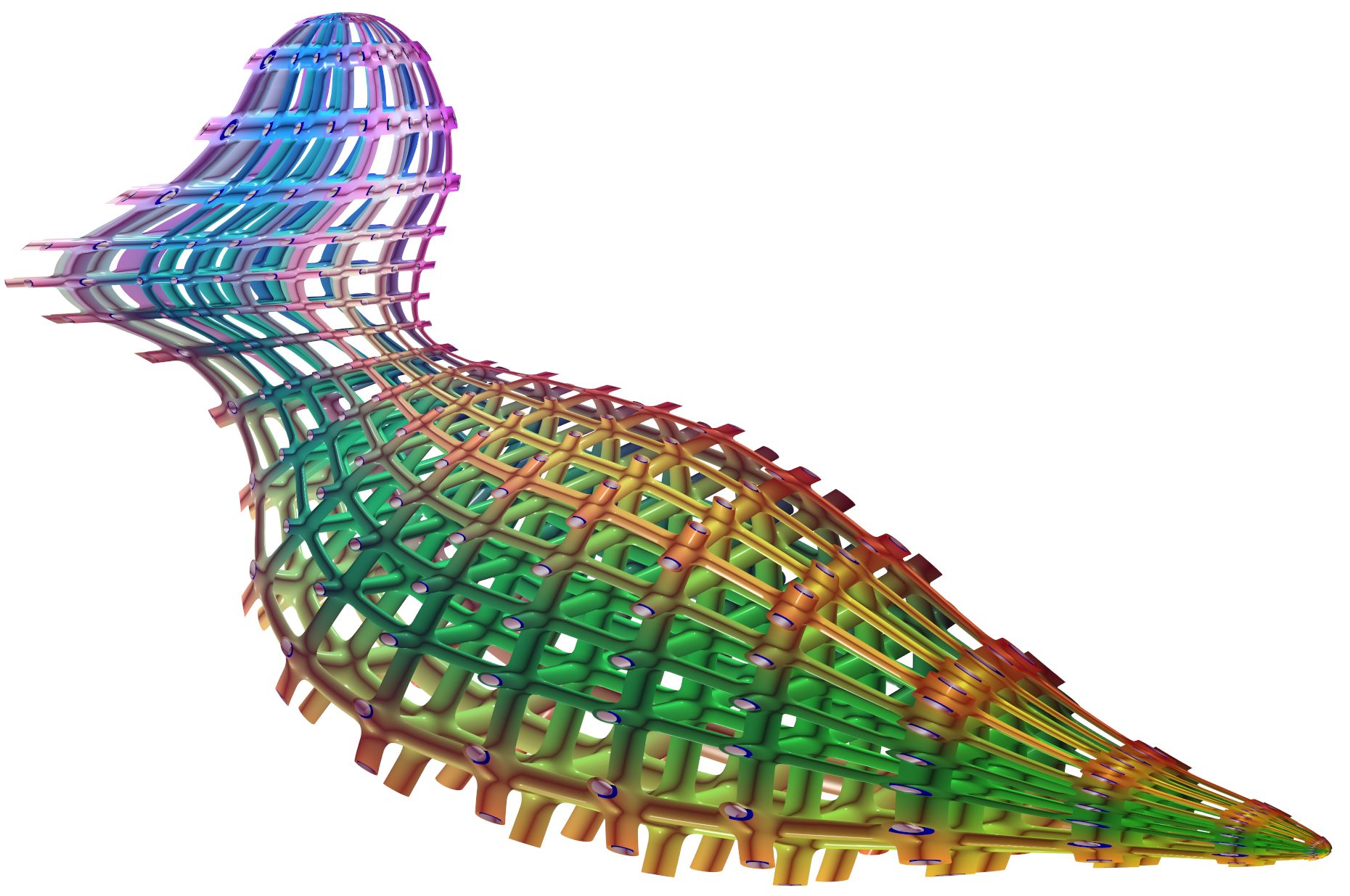}
        \includegraphics[height=1.7in]{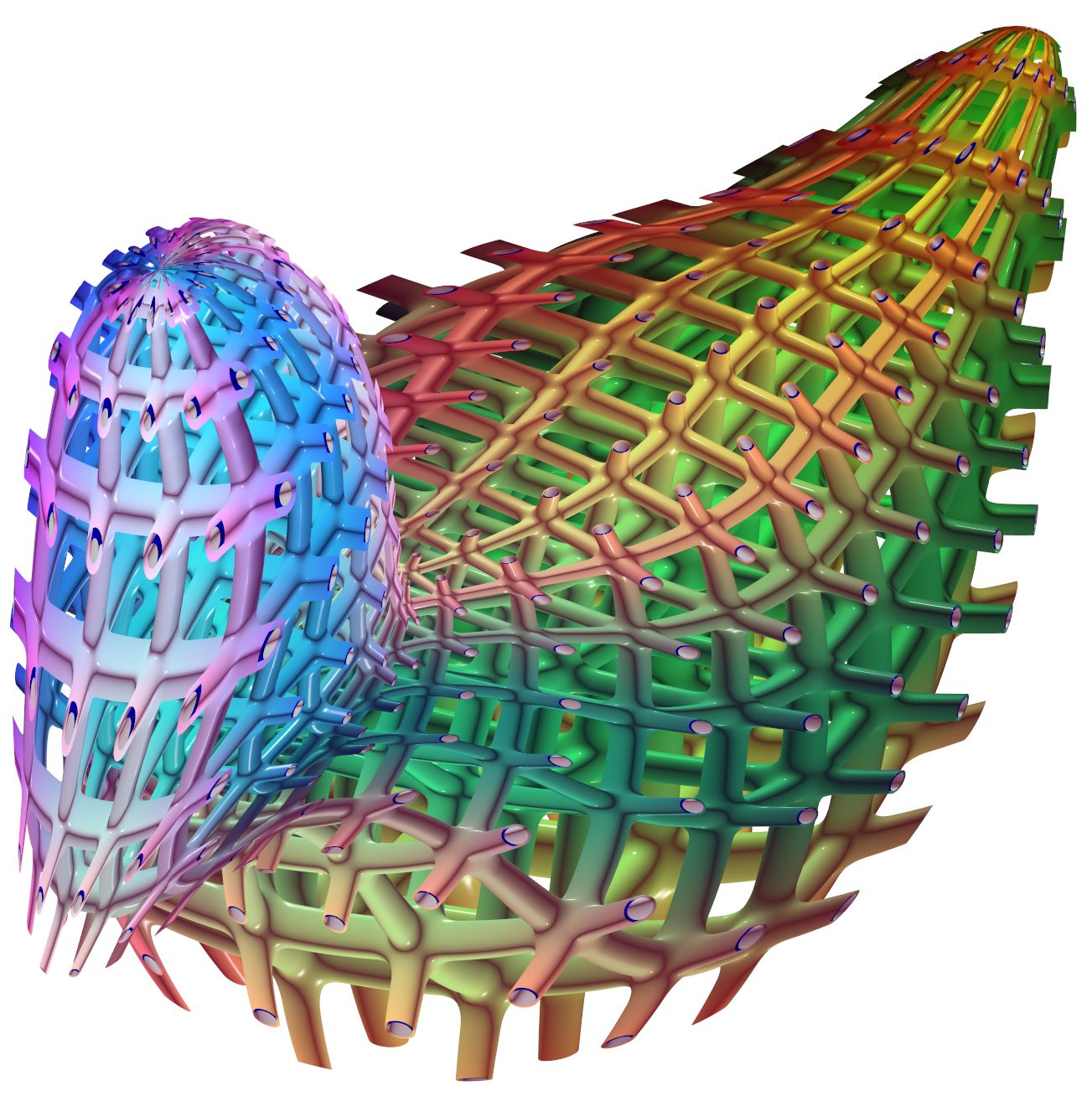}
        \includegraphics[height=1.7in]{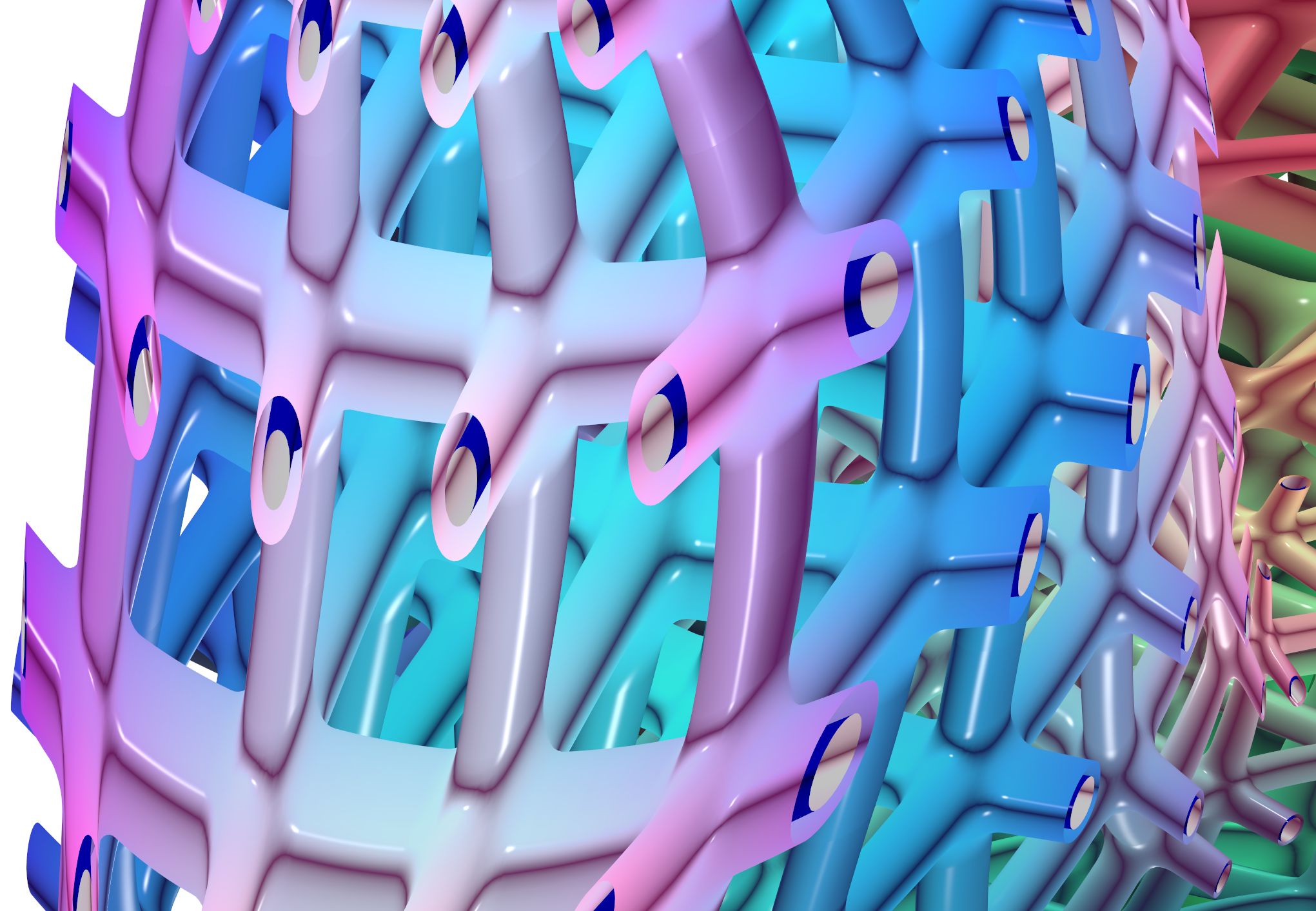} \\
    \end{center}
    \mbox{\vspace{-0.35in}}\\[-0.35in]
    \caption{A heterogeneous duck macro-shape with a variable offset
        of an implicit cross tile.}
    \label{fig-duck-cross-var-offset}
    \mbox{\vspace{-0.5in}}\\[-0.5in]
\end{figure*}

\begin{figure*}
    \begin{center}
        \includegraphics[height=1.9in]{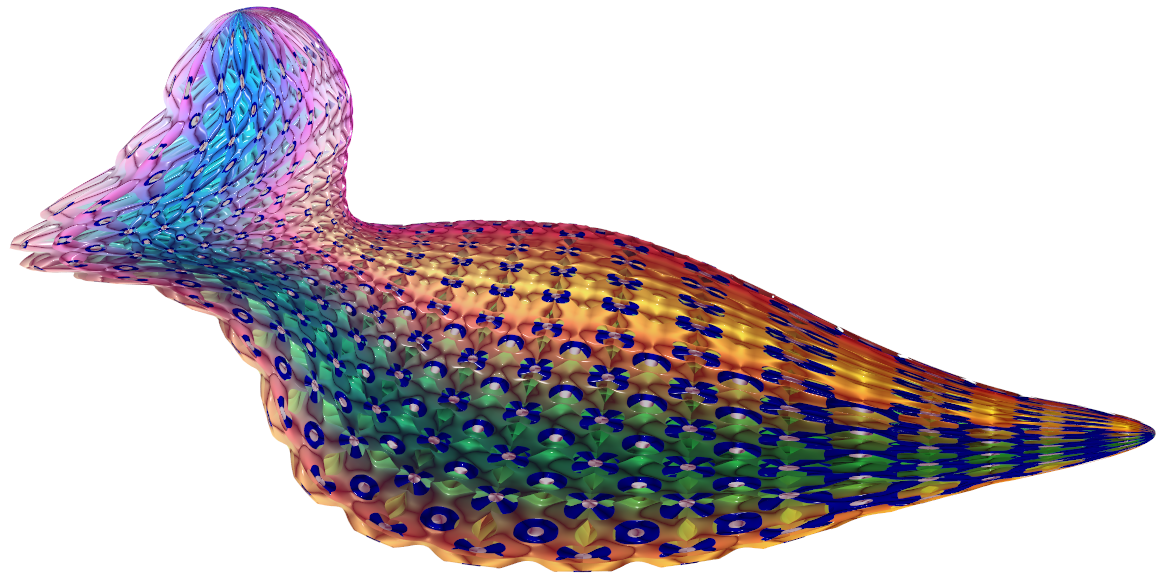}
        \includegraphics[height=1.9in]{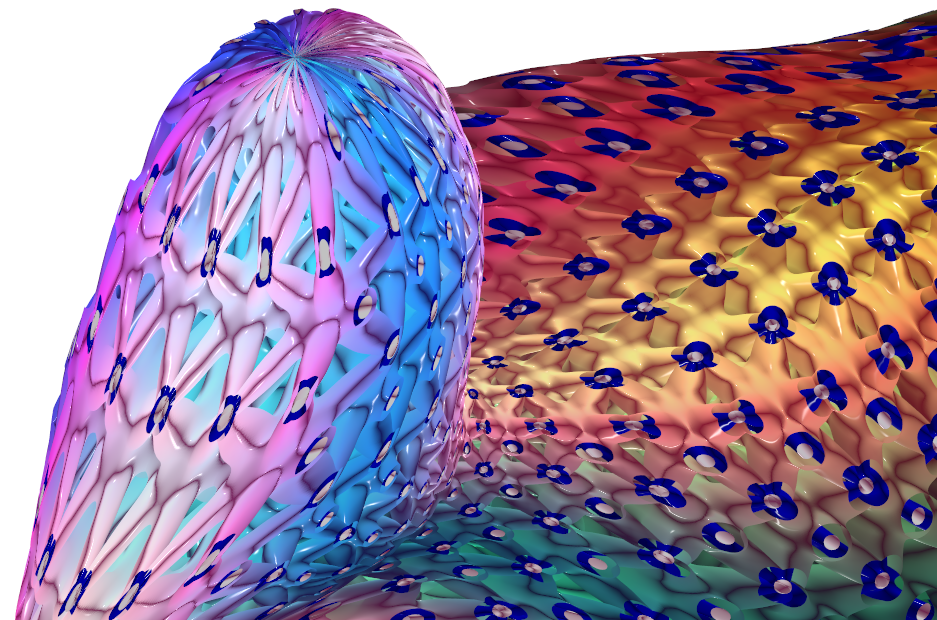} \\
        \begin{picture}(0,0)
            \put( -220,  45){\large (a)}
            \put(  145, 145){\large (b)}
        \end{picture}
    \end{center}
    \mbox{\vspace{-0.50in}}\\[-0.50in]
    \caption{A heterogeneous duck macro-shape with a variable offset
        of an implicit tile from Figure~\ref{fig_imp_tile_example}.}
    \label{fig-duck-fig1-var-offset}
    \mbox{\vspace{-0.5in}}\\[-0.5in]
\end{figure*}
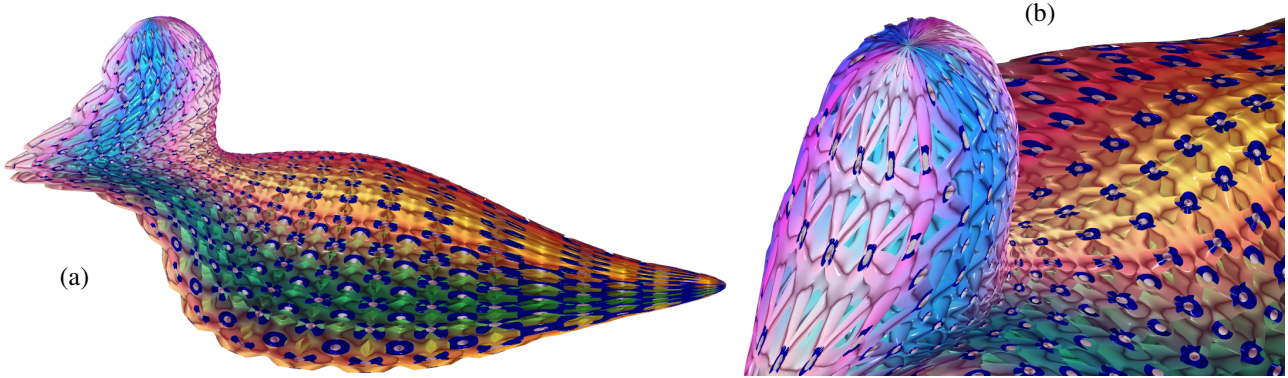

\begin{figure*}
    \begin{center}
        \includegraphics[height=1.7in]{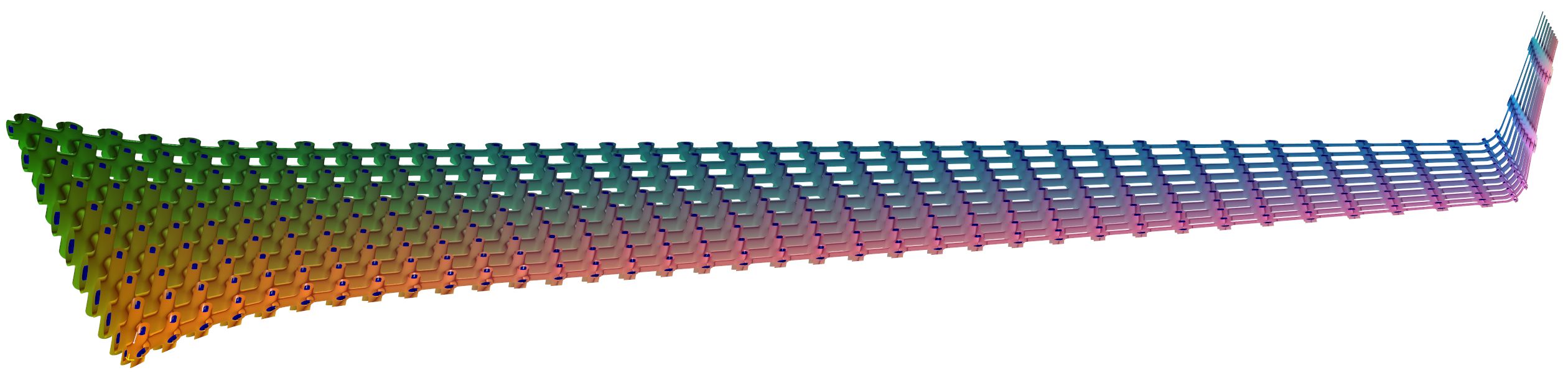}
        \includegraphics[height=2in]{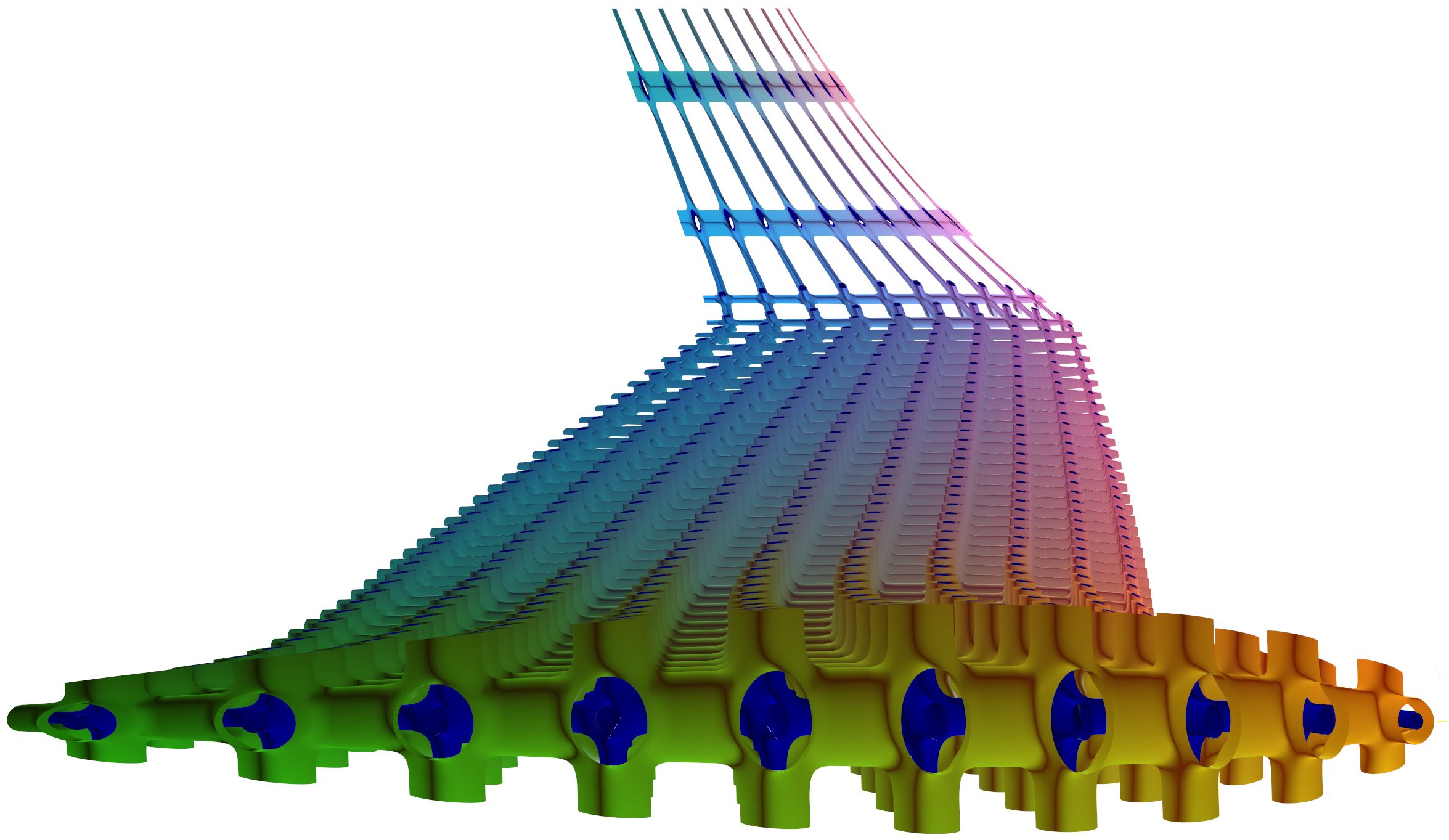}
        \includegraphics[height=2in]{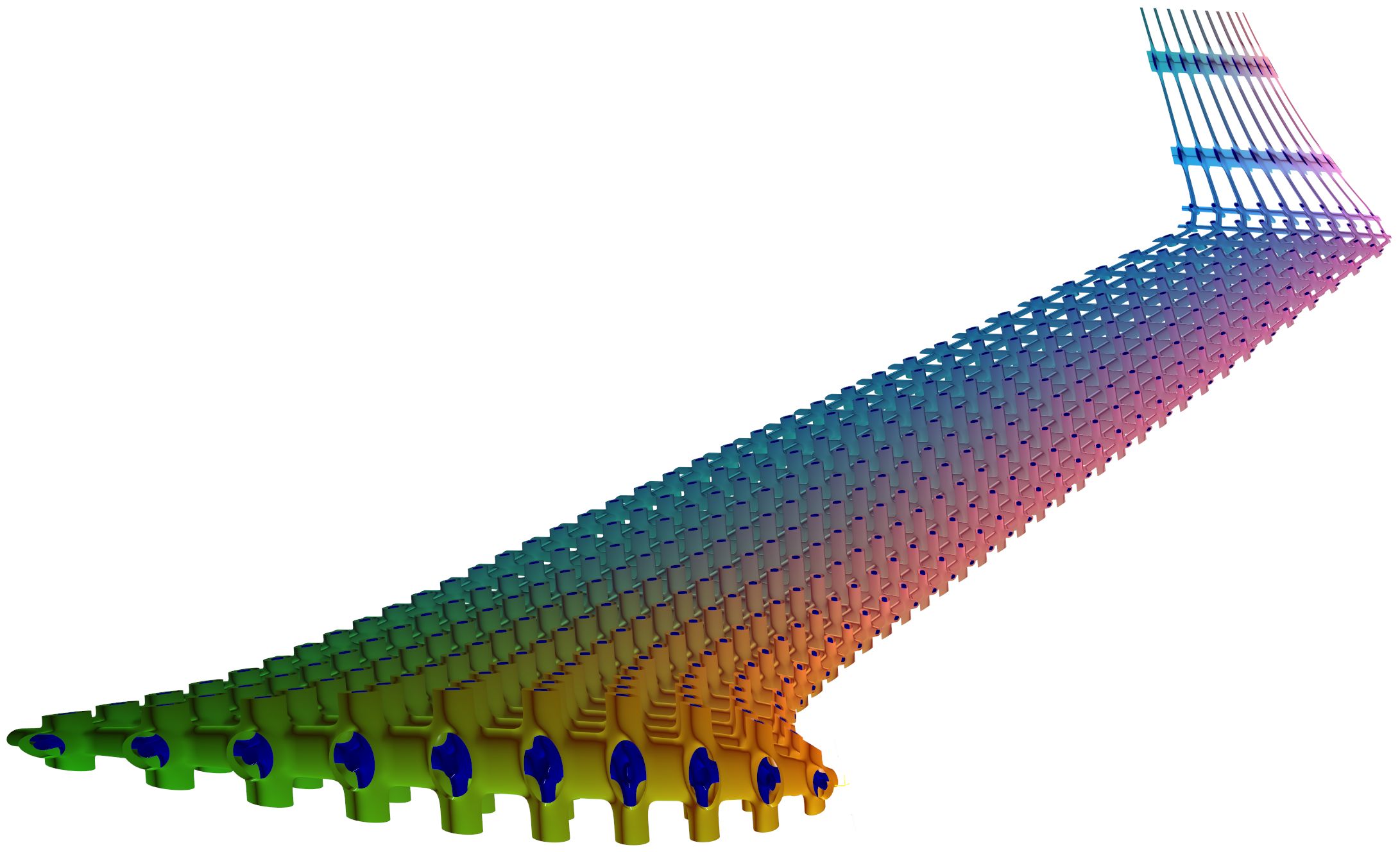} \\
        \begin{picture}(0,0)
            \put(  180, 187){\large (a)}
            \put( -230,  60){\large (b)}
            \put(  180,  40){\large (c)}
        \end{picture}
    \end{center}
    \mbox{\vspace{-0.50in}}\\[-0.50in]
    \caption{A heterogeneous wing macro-shape with a variable offset
        of an implicit cross tile.}
    \label{fig-wing-cross-var-offset}
    \mbox{\vspace{-0.5in}}\\[-0.5in]
\end{figure*}

Figures~\ref{fig-duck-cross-var-offset}
to~\ref{fig-wing-cross-var-offset} show both the original lattice
geometry as well as its variable distance offset of a graded material
(color) duck model (Figures~\ref{fig-duck-cross-var-offset}
and~\ref{fig-duck-fig1-var-offset}) and graded material wing model
(Figure~\ref{fig-wing-cross-var-offset}), with implicit cross tiles in
Figures~\ref{fig-duck-cross-var-offset}
and~\ref{fig-wing-cross-var-offset}, and the tile from
Figure~\ref{fig_imp_tile_example} in
Figure~\ref{fig-duck-fig1-var-offset}. These macro-shape geometries
are trivariates of orders $(3\times3\times3)$ and $(6\times4\times10)$
control points for the duck shape, and of orders $(4\times2\times3)$
and $(4\times2\times10)$ control points for the wing shape, both
defined again in $R^7$ as in Figure~\ref{fig-tiles-cross-imp-diags-axes}.
The computation times of the
marching cube procedure for the offset tiles in the duck shape took
around 25 seconds for constant offsets and close to 40 seconds for
variable distance offsets.  For the wing example, it took 13 seconds
to compute constant offset tiles, and almost 30 seconds for variable
distance offset tiles, on the same i7-12700 machine.

\subsection{Prescribed offsets in Euclidean Lattice space}

\begin{figure}
    \begin{center}
        ~~~~~~~\includegraphics[height=2.15in]{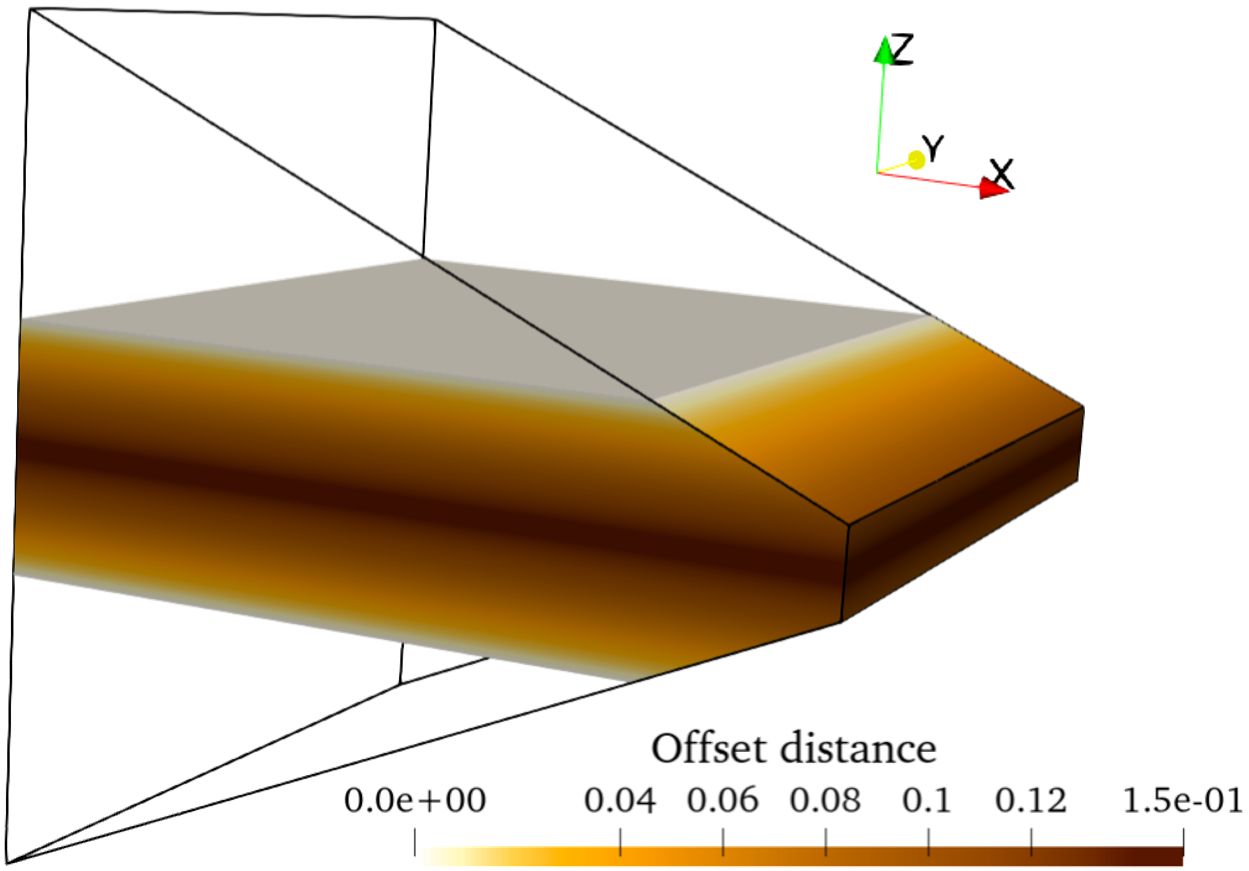} \\
        \begin{picture}(0,0)
            \put(  -110, 162){${\mathcal T}$}
            \put(  -122,  96){\small $z = \frac{1}{2}$}
            \put(  -112,  87){\vector( 1, 0){ 10}}
        \end{picture}
    \end{center}
    \mbox{\vspace{-0.5in}}\\[-0.5in]
    \caption{Linear implicit trivariate $\mathcal{I}(x,y,z)=z-1/2$ mapped
        through a tapered trivariate with prescribed constant offset
        distance $d=0.15$ in Euclidean lattice space on both sides of
        the mid-surface.
        Field colors represent the distance to the offset surfaces in
        Euclidean space.}
    \label{fig-tmps-euc-offset-tapered}
    \mbox{\vspace{-0.5in}}\\[-0.5in]
\end{figure}

In this subsection, we gathered a few examples of generated lattices
that present a prescribed offset in Euclidean lattice space (following
the algorithm introduced in Section~\ref{subsec-prescribed-offsets}).
A first simple example, shown in
Figure~\ref{fig-tmps-euc-offset-tapered}, is designed to highlight the
differences from the approach in
Section~\ref{subsec-create-implicit-offset}.  In this case, we consider
an implicit tile defined by a linear function
$\mathcal{I}(x,y,z)=z-1/2$, mapped through a tapered trivariate
geometry, and with a constant offset distance $d=0.15$ in Euclidean
lattice space.  As can be observed, the lattice presents constant
thickness on both sides of the mid-surface $z=1/2$ and gets skewly
trimmed by the tapered faces of the trivariate.  In
Figure~\ref{fig-tmps-euc-offset-tapered}, as well as in all the other
results in the subsection, the distance to the offset surfaces in
Euclidean space is visualized as a color field on top of the geometry:
The distance is zero (white) on the offset surfaces and maximum
(brown) on the mid-surface.

\begin{figure*}
    \begin{center}
        \includegraphics[height=2.5in]{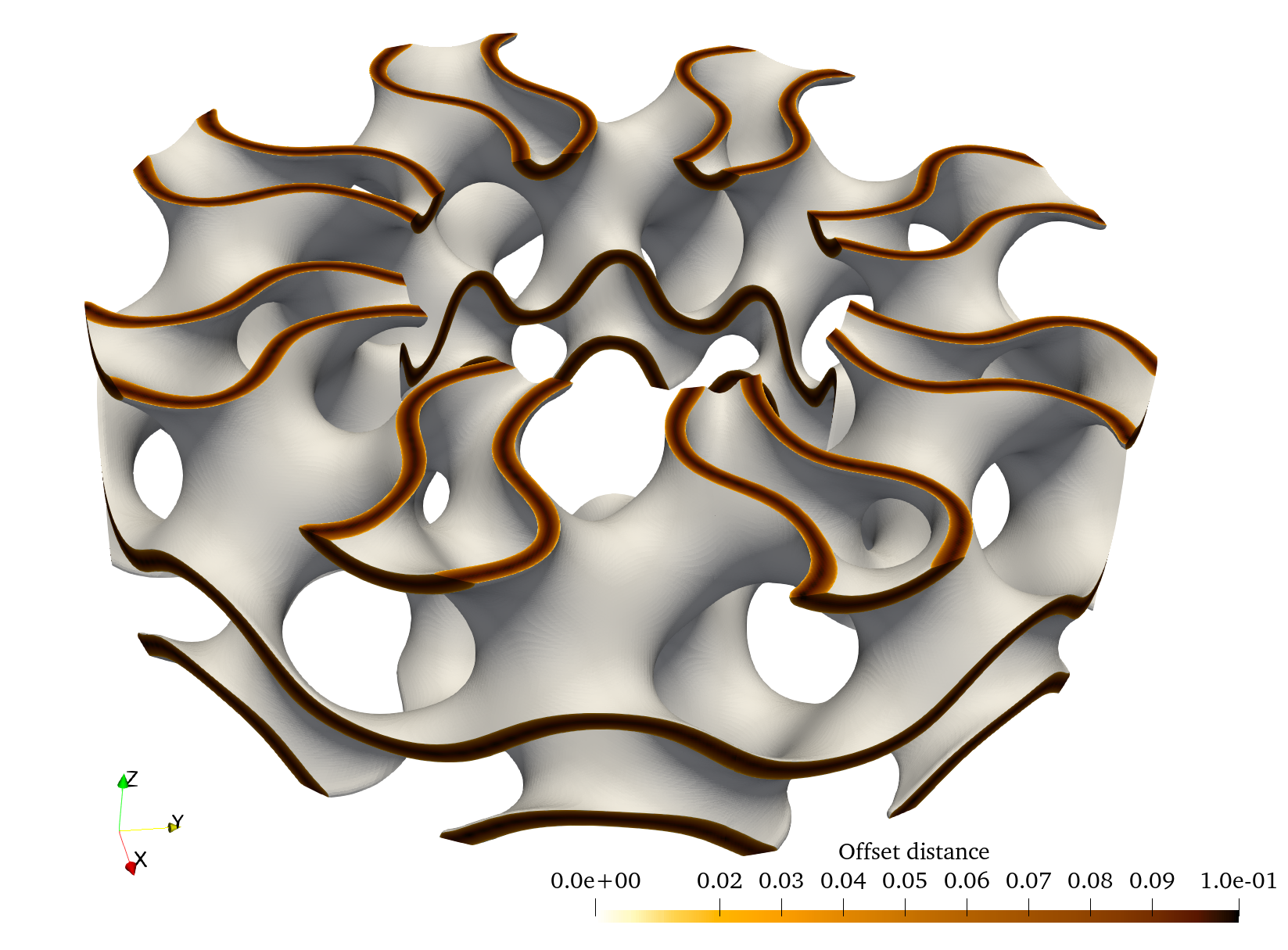}
        \includegraphics[height=2.5in]{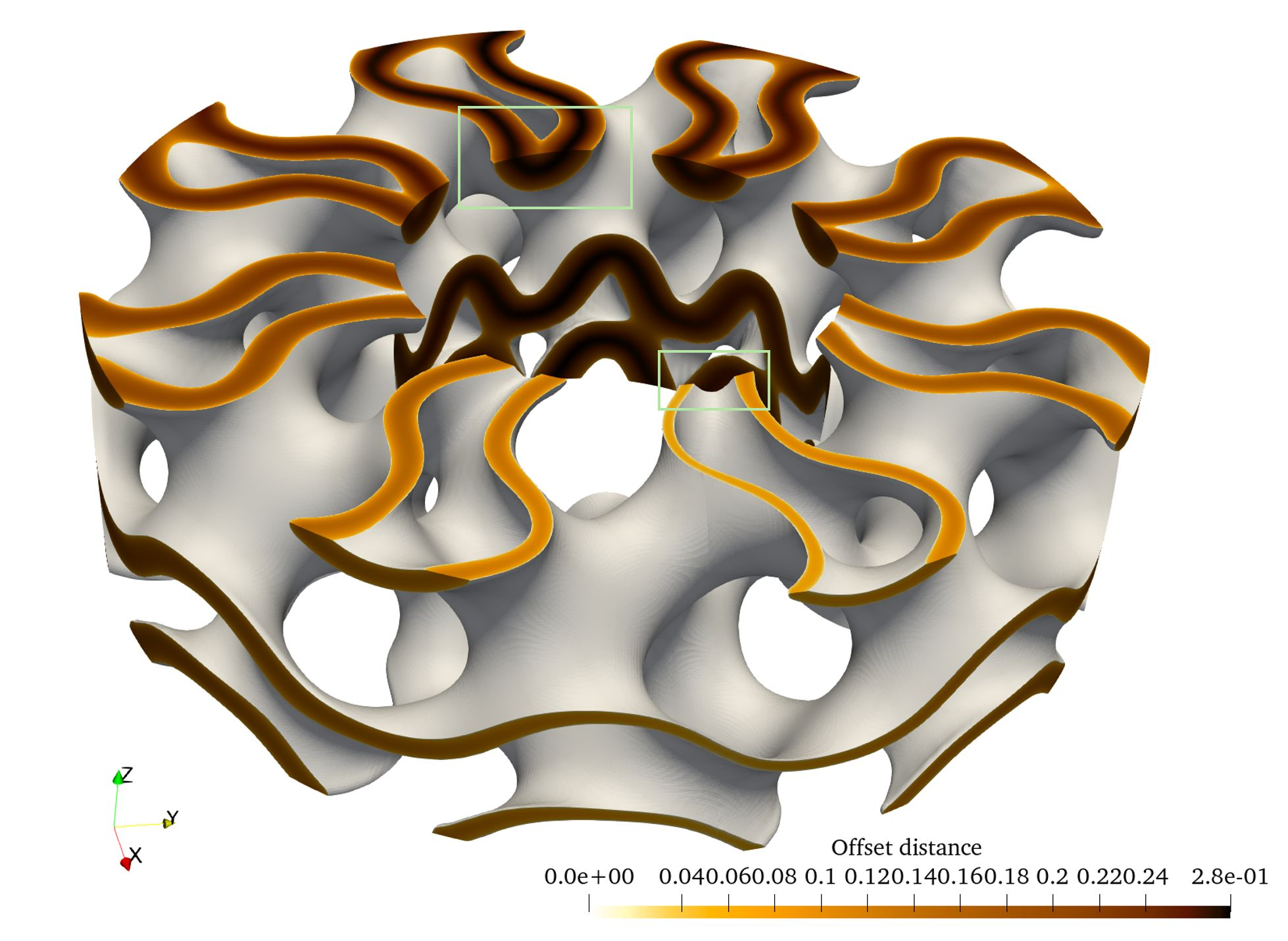} \\[-0.1in]
        \includegraphics[height=2.5in]{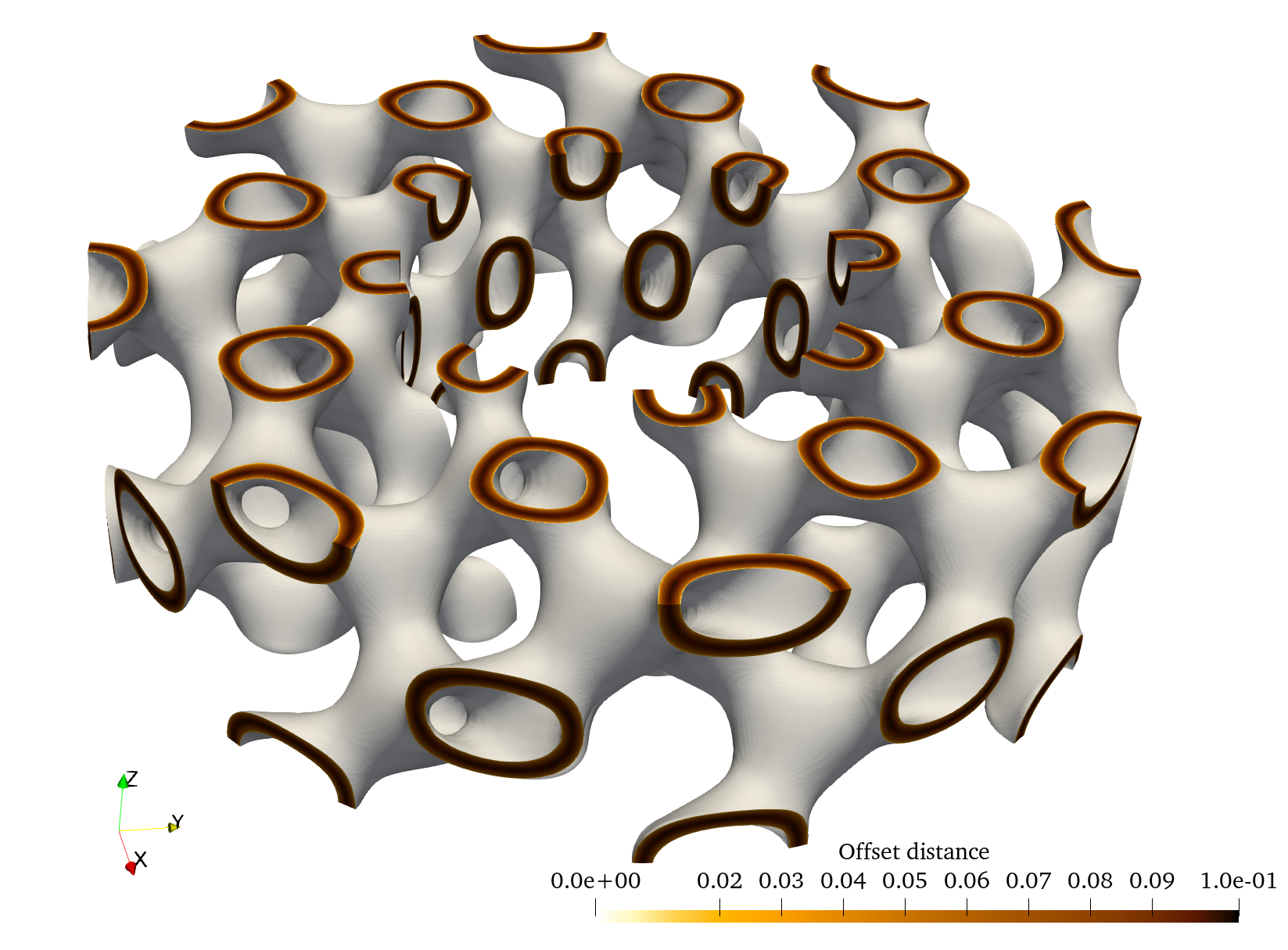}
        \includegraphics[height=2.5in]{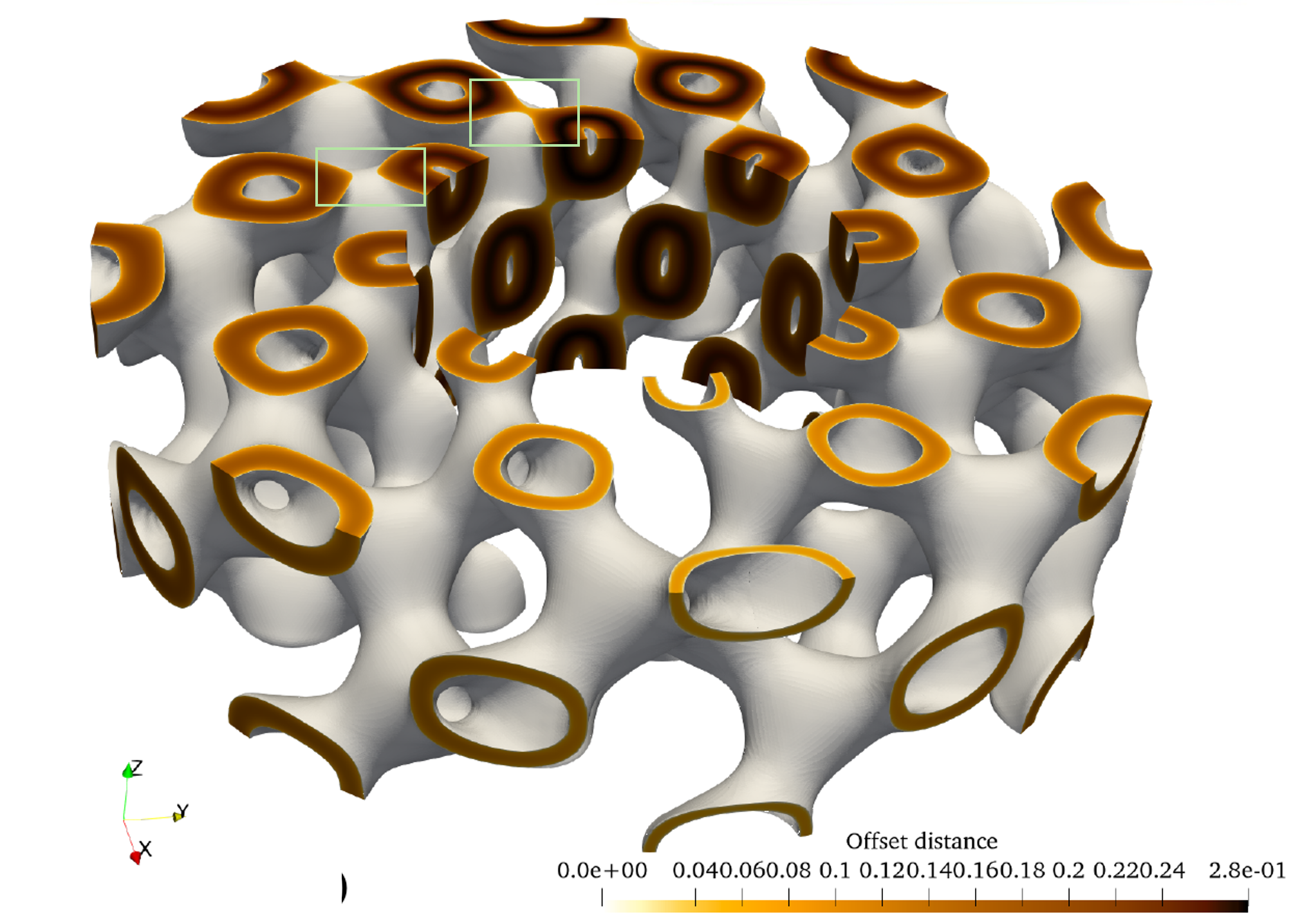} \\
        \begin{picture}(0,0)
            \put( -200, 200){(a)}
            \put(   50, 200){(b)}
            \put( -200, 19){(c)}
            \put(   55, 19){(d)}
        \end{picture}
    \end{center}
    \mbox{\vspace{-0.6in}}\\[-0.6in]
    \caption{Schoen's gyroid (a)-(b) and Scharwz's diamond (c)-(d) implicit trivariates in a torus macro-shape with square cross section and prescribed offset distances in Euclidean space: (a)-(c) constant distance and (b)-(d) variable distance.
        Note in (b) and (d), the smooth topological changes in the offsets, where no self-intersection processsing is required - compare each two marked rectangles for the proper topological change in both small and large offset amounts, in these variable offset examples.
        The brown field colors represent the distance to the offset surfaces in Euclidean space.}
    \label{fig-tmps-euc-offset-torus1}
    \mbox{\vspace{-0.5in}}\\[-0.5in]
\end{figure*}

In Figure~\ref{fig-tmps-euc-offset-torus1}, we present several examples in
which the Schoen's gyroid and Scharwz's diamond implicit tiles
are mapped through a torus with
square cross section that has orders $(4,2,2)$ and $(13,2,2)$ control
points.  In the left part of the figure (i.e.,
Figures~\ref{fig-tmps-euc-offset-torus1}~(a) and~\ref{fig-tmps-euc-offset-torus1}~(c))
the tiles are generated with a
constant offset distance $d=0.1$ symmetrically on both sides of the
mid-surface, prescribed in Euclidean lattice space; while in the right
part (Figures~\ref{fig-tmps-euc-offset-torus1}~(b) and~\ref{fig-tmps-euc-offset-torus1}~(d))
the offset varies linearly and
periodically along the circumferential parametric direction, from
$d_{\text{min}}=0.1$ up to $d_{\text{max}}=0.28$, on both sides of the
mid-surface.
In Figures~\ref{fig-tmps-euc-offset-torus1}~(b) and~(d),
the topological changes are smoothly accomodated for the different
offset amounts.  No self-intersection processsing is needed - examine
the two marked rectangles in (b) and in (d), having different
topologies in the small and the large offsets.

In all cases, the implicit offsets ${\mathbf I}_{+e}$ and ${\mathbf I}_{-e}$
were approximated as one single \Bspline{} trivariate each containing
all the tiles in the macro-shape.  These \Bspline{} approximations
have orders $(3,3,3)$ and $(259,35,35)$ control points in $\Reals^1$,
with $C^1$ continuity at the tiles' interfaces.

\begin{figure*}
    \begin{center}
        \mbox{\vspace{-0.15in}}\\[-0.15in]
        \includegraphics[height=1.75in]{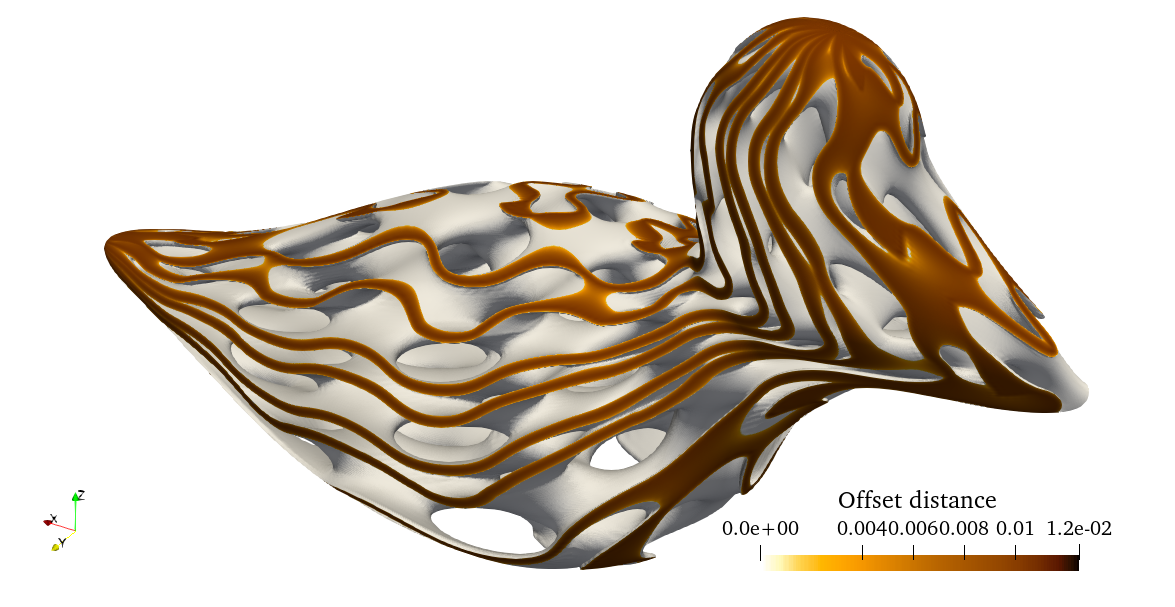}
        \includegraphics[height=1.75in]{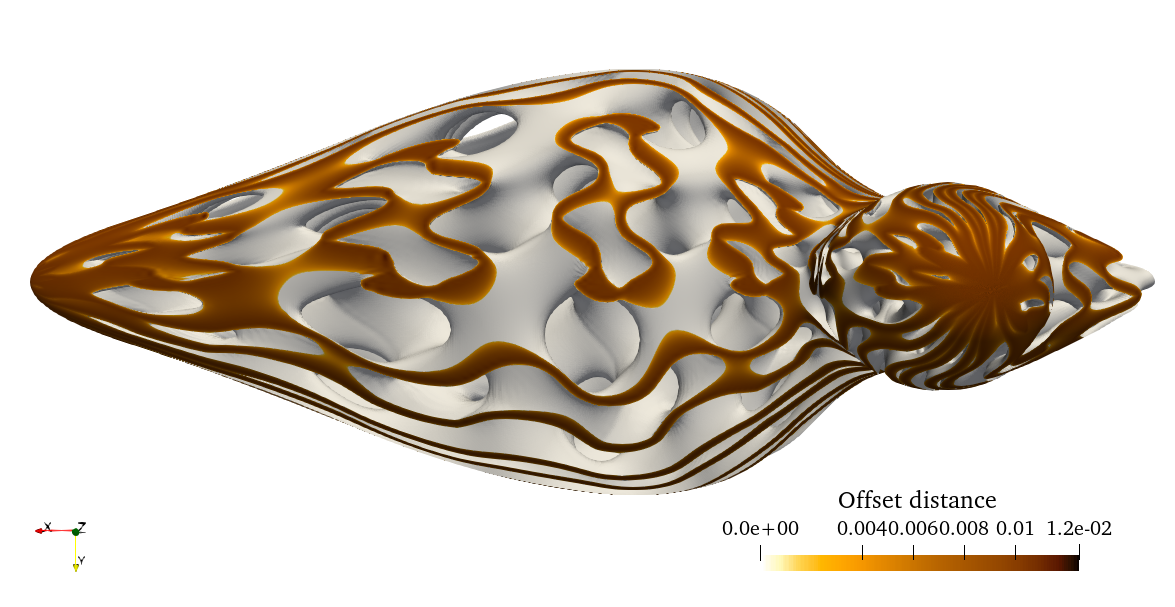}\\[-0.13in]
        \includegraphics[height=1.75in]{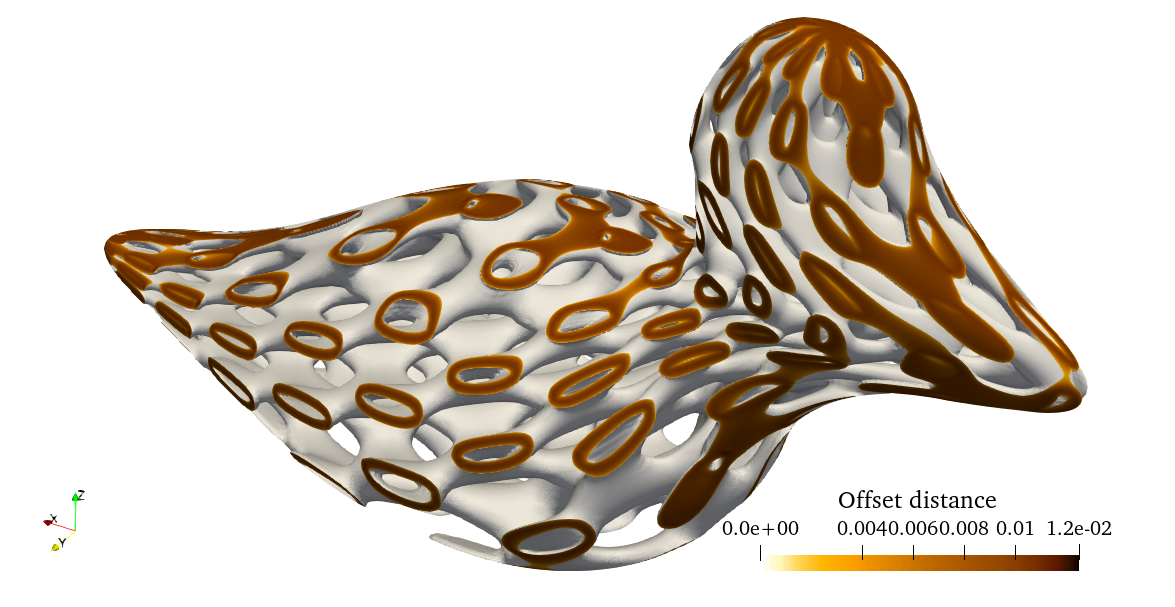}
        \includegraphics[height=1.75in]{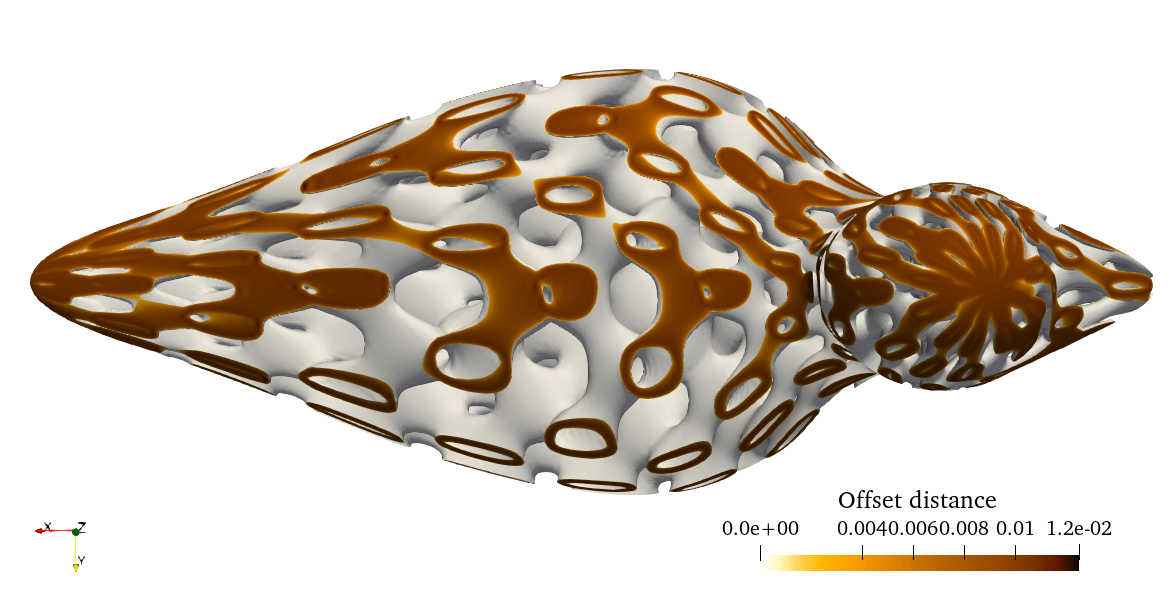}\\
        \begin{picture}(0,0)
            \put( -220, 155){(a)}
            \put(   30, 155){(b)}
            \put( -220, 25){(c)}
            \put(   30, 25){(d)}
        \end{picture}
    \end{center}
    \mbox{\vspace{-0.65in}}\\[-0.65in]
    \caption{Schoen's gyroid (a)-(b) and Scharwz's diamond (c)-(d) implicit trivariates in a duck macro-shape with constant offset distances in Euclidean space: (a)-(c) perspective and (b)-(d) top view.
        The brown Field colors represent the distance to the offset surfaces in Euclidean space.}
    \label{fig-tmps-euc-offset-duck1}
\end{figure*}

In a similar way, in Figure~\ref{fig-tmps-euc-offset-duck1},
$(18,5,4)$ implicit tiles, with the same analytic implicit definitions
as before, were mapped through a duck geometry that has orders
$(3,3,3)$ and $(6,4,10)$ control points.  A constant offset distance
$d=0.01$ on both sides of the mid-surface was considered in all the
cases and the offsets ${\mathbf I}_{+e}$ and ${\mathbf I}_{-e}$ were
approximated as one single \Bspline{} trivariate each with orders
$(3,3,3)$ and $(259,35,35)$ control points in $\Reals^1$, with $C^1$
continuity at the tiles' interfaces.  As it can be observed in the
figures, the different size of the duck's \Bezier{} elements, combined
with the constant distance offset of the tiles, induces severe
variations in the tiles' topology, with some of them (as the ones near
the tail and head) almost solid.

All the examples in this subsection were computed using a single
processor in an Apple M2 Max chip with 64 GB of DRAM memory.  The
generation of the implicit offsets ${\mathbf I}_{+e}$ and ${\mathbf I}_{-e}$
and their approximation with \Bspline{} trivariates took between
$0.41$ and $1.24$ seconds for the cases in
Figure~\ref{fig-tmps-euc-offset-torus1}, and between $3.05$ and $8.73$
seconds for the ones in Figure~\ref{fig-tmps-euc-offset-duck1}.

\subsection{Simulation}

\begin{figure*}
    \begin{center}
    \mbox{\vspace{-0.3in}}\\[-0.3in]
        \includegraphics[width=0.45\textwidth]{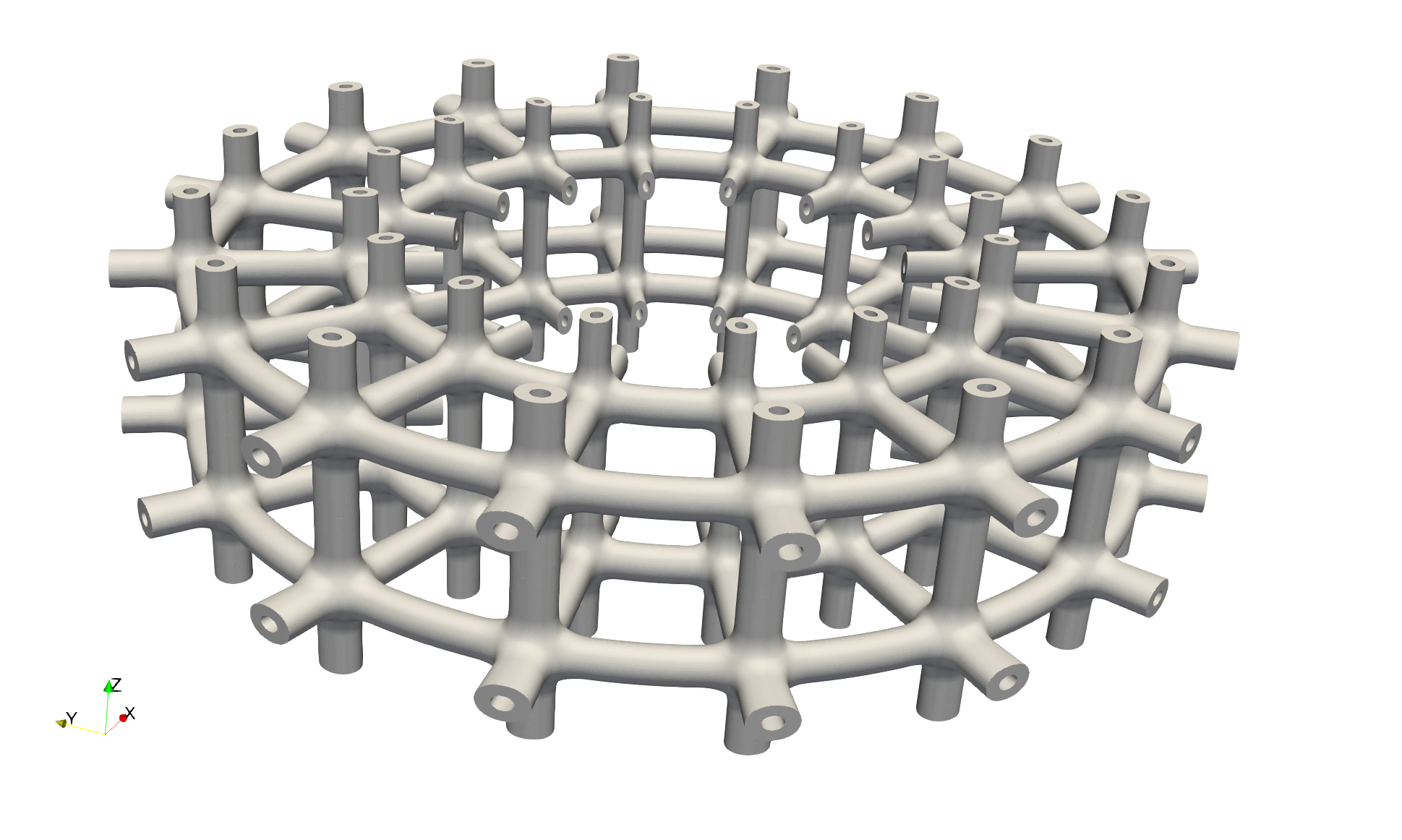}
        \includegraphics[width=0.45\textwidth]{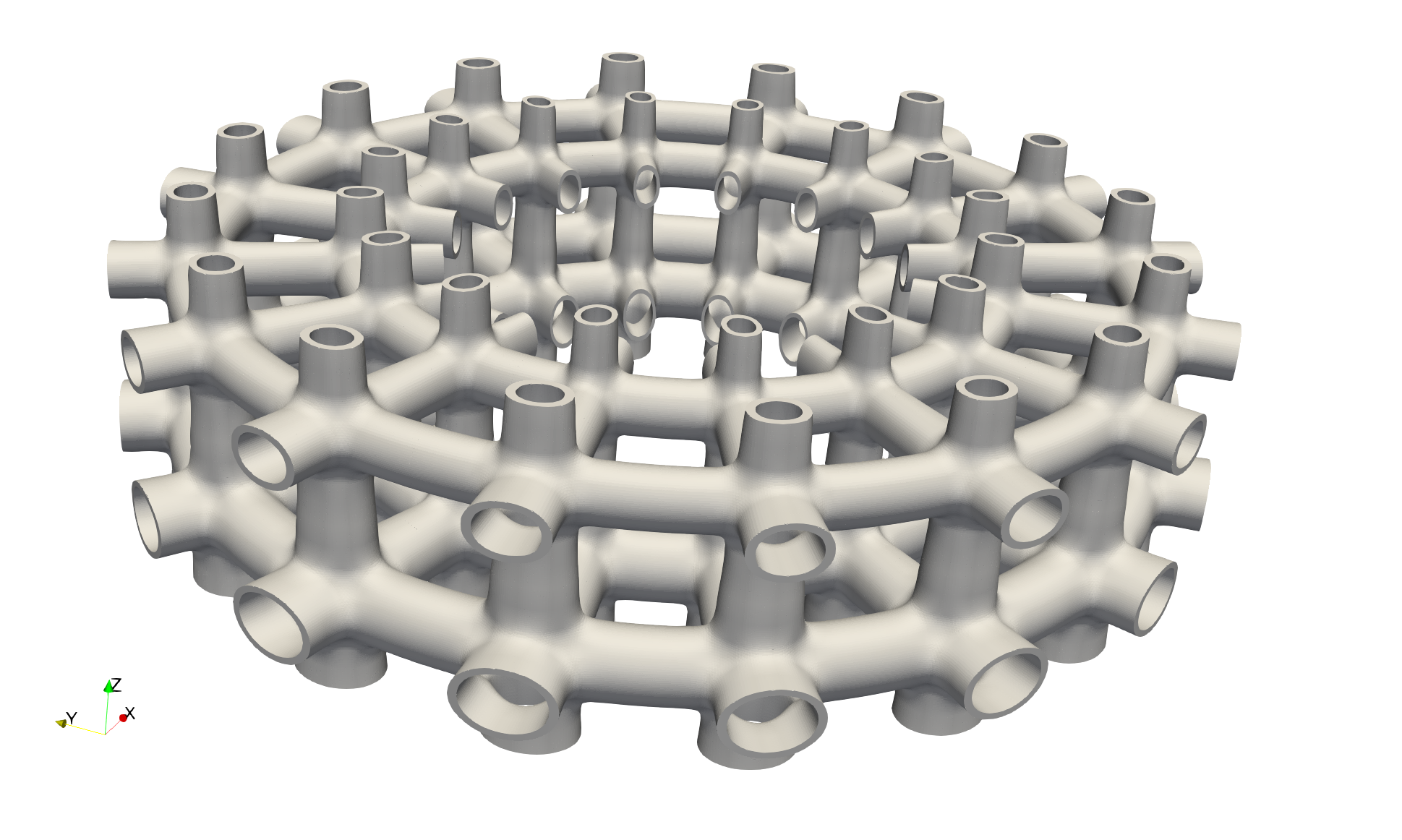}\\[-0.15in]
        \includegraphics[width=0.45\textwidth]{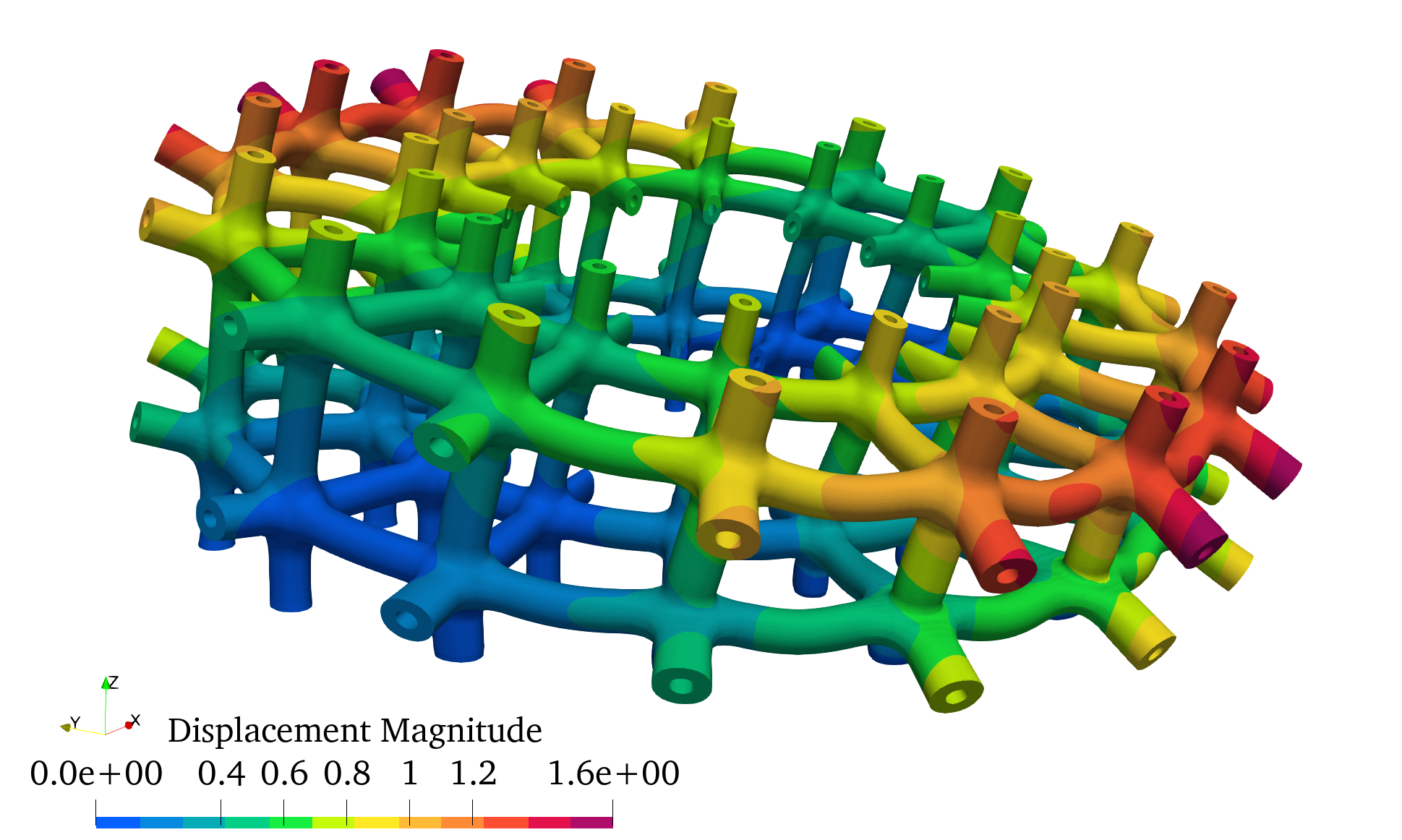}
        \includegraphics[width=0.45\textwidth]{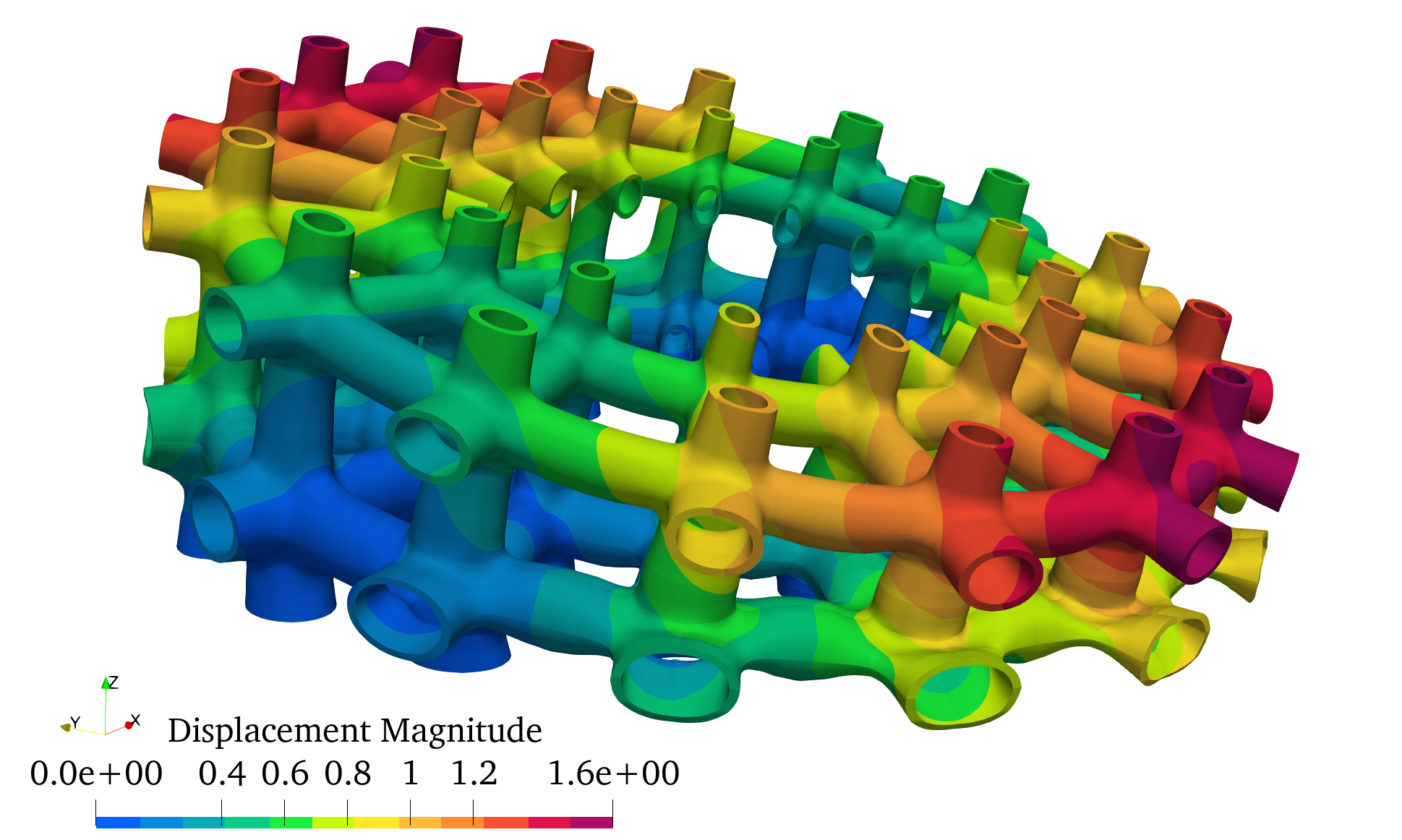} \\
        \begin{picture}(0,0)
            \put(  -37, 167){(a)}
            \put(  200, 170){(b)}
            \put(  -40, 25){(c)}
            \put(  200, 25){(d)}
        \end{picture}
    \end{center}
    \mbox{\vspace{-0.50in}}\\[-0.50in]
    \caption{Simulation of TPMS implicit trivariates, in a torus with square cross section:
        In (a)-(b) two different sets of tiles are considered.
        Their bottom faces are fixed and bending distributed loads are applied.
        Their deformed configurations with elastic deformation fields are depicted in (c)-(d), respectively.}
    \label{fig-simulation_1}
    \mbox{\vspace{-0.5in}}\\[-0.5in]
\end{figure*}

\begin{figure*}
    \begin{center}
        \mbox{\vspace{-0.12in}}\\[-0.12in]
        \includegraphics[width=0.99\textwidth]{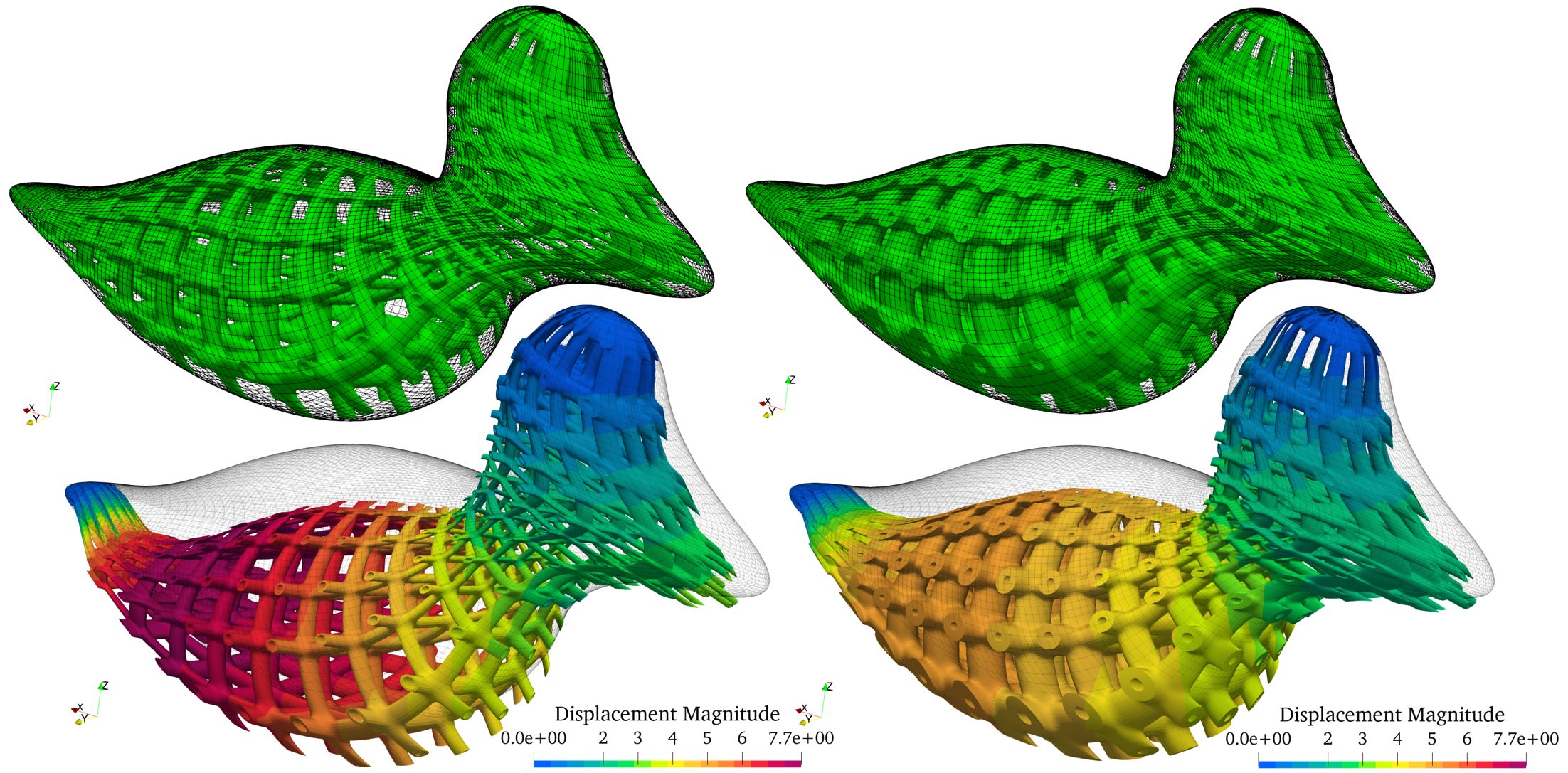} \\
        \begin{picture}(0,0)
            \put( -235, 215){(a)}
            \put(   10, 220){(b)}
            \put( -232,  50){(c)}
            \put(   10,  50){(d)}
        \end{picture}
    \end{center}
    \mbox{\vspace{-0.5in}}\\[-0.5in]
    \caption{Simulation of TPMS implicit trivariates in a duck macro-shape:
        (a) and (b) show the implicit geometries immersed in the solution
        background mesh and (c) and (d) presents the deformed geometries
        with elastic deformation field.  Two different cross tiles were
        considered: thinner in (a) and (c), and thicker in (b) and (d).  For
        the simulations in (c) and (d), the external faces of the duck's
        tale and the top part of the head are fixed, and the same vertical
        gravity load is applied in the whole body.}
    \label{fig-simulation_3}
    \mbox{\vspace{-0.5in}}\\[-0.5in]
\end{figure*}

To illustrate the possibility of running finite element simulations
directly using the constructed geometries, we present two examples
(Figures~\ref{fig-simulation_1} and~\ref{fig-simulation_3}) following
the methodology proposed in Section~\ref{sec-analysis-implicit}.  In
both cases, we solve a linear elasticity problem
(see {\MyungSoo Appendix}~\ref{sec-appendix}), setting the material mechanical properties
as: Young's modulus $E=1$ and Poisson's ratio $\nu=0.3$.  In all
cases, just part of the boundary is constrained\footnote{As detailed
    in~\ref{sec-appendix}, and for the sake of simplicity, Dirichlet
    boundary conditions are only applied on a subset $\Gamma_D$ of the
    boundary that is also part of the boundary of the grid $\mathcal{G}$,
    i.e., $\Gamma_D\subset\partial\Omega\cap\partial\mathcal{G}$.}, the
rest is traction free (homogeneous Neumann condition), and a
distributed load is applied everywhere in the model.

The first test case corresponds to TPMS cross tiles as implicit
trivariates that are mapped through a torus with square cross section.
$(16,2,2)$ tiles were placed in the parametric domain of the torus,
where every cross tile trivariate has orders $(3,3,3)$ and $(10,10,10)$
control points in $\Reals^1$, and the macro-shape is of order
$(4,2,2)$ and has $(13,2,2)$ control points.  For the PDE
discretization we considered a background tensor-product grid with
$(128,16,16)$ elements, coincident with the arrangement of all the
implicits, and trilinear basis functions.

The displacement of the bottom surface of the tori is
blocked, while a bending distributed load $(0,0,\alpha\,y)$ is
applied, with $\alpha=2\cdot10^{-2}$ for the case in
Figure~\ref{fig-simulation_1}~(c), and $\alpha=8.5\cdot10^{-2}$ for
Figure~\ref{fig-simulation_1}~(d), the tori being centered at the origin
and the $z$ axis pointing upwards.  While both designs present a
similar amount of material (volumes are $20.82$ and $21.25$,
respectively) and level of deformation (see the displacement magnitude
fields in the corresponding figures), the geometry in
Figure~\ref{fig-simulation_1}~(b) results in a much stiffer design,
withstanding a load more than $4$ times higher than the one in
Figure~\ref{fig-simulation_1}~(a).

In Figure~\ref{fig-simulation_3} another simulation with a more
complex design is shown.  In this case, we consider cross tile implicit
trivariates mapped through a duck geometry that has orders $(3,3,3)$
and $(6,4,10)$ control points.  $(18,5,4)$ tiles were placed in the
parametric domain of the duck, where each cross tile trivariate has
orders $(3,3,3)$ and $(10,10,10)$ control points in $\Reals^1$.  The
tiles were created with variable distance offsets (following the
algorithm in Section~\ref{subsec-create-implicit-offset}),
and two different levels of thickness were considered
(see Figures~\ref{fig-simulation_3}~(a) and~\ref{fig-simulation_3}~(b)).
The elastic solution
was discretized using a background tensor-product grid with
$(144,40,32)$ elements, coincident with the arrangement of all the
implicits, and
trilinear basis functions (this grid is shown in
Figure~\ref{fig-simulation_3}).  In this case, the elastic
displacement is blocked for the nodes of the head and tail tiles that
belong to the boundary of the macro-shape, while a gravity load
$(0,0,-0.20)$ is applied, inducing a bending behavior in the duck as
observed in Figures~\ref{fig-simulation_3}~(c)
and~\ref{fig-simulation_3}~(d).
As can be expected, the bulkier design (Figure~\ref{fig-simulation_3}~(d))
presents a stiffer behavior.

All simulations were run using a single processor in an Apple
M2 Max chip with 64 GB of DRAM memory, using the FEniCSx open source
computing platform~\cite{dolfinx,basix,ufl}, together with extra
capabilities developed for dealing with unfitted discretizations that
were implemented on top of the open source projects
\texttt{algoim}~\cite{algoim} and
\texttt{customquad}~\cite{customquad}.  The analysis times of these
examples, without considering the time required for generating the
output visualization files, were $11.4$, $14.5$, $89.7$, and $143.8$ seconds,
for the cases in Figures~\ref{fig-simulation_1}~(c), \ref{fig-simulation_1}~(d),
\ref{fig-simulation_3}~(c), and \ref{fig-simulation_3}~(d), respectively



\section{Conclusions and Future Work}
\label{sec-conclude}

In this work, we have portrayed schemes to build controlled, constant
and variable, offsets for implicit forms, in both the
Euclidean space and even in the lattice space after the mapping
through the macro-shape.  Further, the use of immersive method to
analyze the created 3D geometries was further presented, enabling
a direct closed loop between design and analysis.

In the conventional geometric and solid modeling systems, a proper
trimming of offset self-intersections has always been one of the main
technical issues for a robust implementation of the underlying
systems. Yet, and as demonstrated in
Section~\ref{sec-results}, the use of implicit forms can alleviate
this self-intersection difficulty.  Though limited in resolution, the
representation of distance fields in voxelized spaces has been
employed as an effective solution in
practice~\cite{McMains2014,Kobbelt,CLWang}.  In this work, we have
presented a \Bspline{}-based approach that can extend the discrete
representation of distance fields to a generic functional modeling of
implicit fields for the synthesis and analysis of heterogeneous
microstructures. The current approach introduces approximation errors
in different levels and from various sources, due to the result of
functional compositions over heterogeneous forms.  In future work, we
would like to explore the direction of systematically controlling the
sampling of (procedurally defined) implicit functions and reducing the
error to arbitrary precisions, suitable for specific applications
under consideration.

We also plan to extend the presented simulation capabilities
to the case of implicits that present heterogeneous material
properties, and explore the possibility of leveraging on domain
decomposition and reduced order modeling techniques to speed up the
simulations for a high number of implicit tiles by exploiting the
intrinsic similarities among lattices, in a similar way as in the
recent work~\cite{Hirschler2024}.  Other physics problems of interest,
namely fluid dynamics, heat transfer, or waves propagation, will be
also considered.

{\Gershon During offsets of implicits, the topology might change (as
seen, for example, in Figure~\ref{fig-tmps-euc-offset-torus1}). A
possible direction to explore here is the detection and possibly the
elimination of these topological changes, if so desired.

All the implicit modeling implementation parts
(Section~\ref{subsec-create-implicit-tiles}) including variable
distance offsets of implicit
(Section~\ref{subsec-create-implicit-offset}) is now part of the Irit
geometric modeling kernel~\cite{IRIT24} and is freely available at the
source code level for non commercial use.  We hope to publically
release the other portions of the implementation in the near future}.


\section*{Acknowledgments}

This research was supported in part by the ISRAEL SCIENCE FOUNDATION
(grant No. 1817/24) and in part by the National Research Foundation of
Korea (NRF) grant funded by the Korea government (MSIT)
(RS-2024-00352439).  Pablo Antolin acknowledges the support of
the Swiss National Science Foundation through the project FLASh (No 214987).
We would also like to thank anonymous reviewers
for their invaluable comments.

\typeout{}


\bibliographystyle{eg-alpha-doi}
\bibliography{MSImplicitOffset}

\newcommand{\etalchar}[1]{$^{#1}$}
\begin{thebibliography}{\uppercase{dPVvB{\etalchar{*}}23}}

\bibitem[AES{\etalchar{*}}19]{Abueidda2019}
\textsc{Abueidda D.~W., Elhebeary M., Shiang C.-S.~A., Pang S., {Abu Al-Rub}
  R.~K., Jasiuk I.~M.}:
\newblock Mechanical properties of 3d printed polymeric gyroid cellular
  structures: Experimental and finite element study.
\newblock \emph{Materials \& Design 165} (2019), 107597.

\bibitem[ALO{\etalchar{*}}14]{ufl}
\textsc{Aln\ae{}s M.~S., Logg A., \O{}lgaard K.~B., Rognes M.~E., Wells G.~N.}:
\newblock Unified form language: A domain-specific language for weak
  formulations of partial differential equations.
\newblock \emph{ACM Transactions on Mathematical Software 40}, 2 (2014), 1--23.

\bibitem[BCH{\etalchar{*}}15]{cutfem}
\textsc{Burman E., Claus S., Hansbo P., Larson M.~G., Massing A.}:
\newblock Cutfem: Discretizing geometry and partial differential equations.
\newblock \emph{International Journal for Numerical Methods in Engineering
  104}, 7 (2015), 472--501.

\bibitem[BDD{\etalchar{*}}23]{dolfinx}
\textsc{Baratta I., Dean J., Dokken J.~S., Habera M., Hale J., Richardson C.,
  Rognes M., Scroggs M., Sime N., Wells G.}:
\newblock Dolfinx: The next generation fenics problem solving environment.
\newblock Preprint, doi.org/10.5281/zenodo.10447665, 12 2023.

\bibitem[BFK92]{Barnhill}
\textsc{Barnhill R.~E., Frost T.~M., Kersey S.~N.}:
\newblock \emph{Self-Intersections and Offset Surfaces}.
\newblock 1992, pp.~35--44.

\bibitem[BHL{\etalchar{*}}16]{burman2016}
\textsc{Burman E., Hansbo P., Larson M.~G., Massing A., Zahedi S.}:
\newblock Full gradient stabilized cut finite element methods for surface
  partial differential equations.
\newblock \emph{Computer Methods in Applied Mechanics and Engineering 310}
  (2016), 278--296.

\bibitem[Bli82]{Blinn1982}
\textsc{Blinn J.~F.}:
\newblock A generalization of algebraic surface drawing.
\newblock \emph{ACM Trans.~on Graphics 1}, 3 (1982), 235–256.

\bibitem[Blo95]{Bloomenthal1995}
\textsc{Bloomenthal J.}:
\newblock \emph{Introduction to Implicit Surfaces}.
\newblock Morgan Kaufmann, 1995.

\bibitem[BPV20]{buffa2020}
\textsc{Buffa A., Puppi R., V\'{a}zquez R.}:
\newblock A minimal stabilization procedure for isogeometric methods on trimmed
  geometries.
\newblock \emph{SIAM Journal on Numerical Analysis 58}, 5 (2020), 2711--2735.

\bibitem[BVM18]{badia2018}
\textsc{Badia S., Verdugo F., Martín A.~F.}:
\newblock The aggregated unfitted finite element method for elliptic problems.
\newblock \emph{Computer Methods in Applied Mechanics and Engineering 336}
  (2018), 533--553.

\bibitem[CRE01]{Cohen2001}
\textsc{Cohen E., Riesenfeld R., Elber G.}:
\newblock \emph{Geometric Modeling with Splines: An Introduction}.
\newblock AK Peters, Wellesley, MA, 2001.

\bibitem[dPVvB{\etalchar{*}}23]{deprenter2023}
\textsc{de~Prenter F., Verhoosel C., van Brummelen E., Larson M., S. B.}:
\newblock Stability and conditioning of immersed finite element methods:
  analysis and remedies.
\newblock \emph{Archives of Computational Methods in Engineering 30} (2023),
  3617--3656.

\bibitem[DPYR08]{duester2008}
\textsc{D\"uster A., Parvizian J., Yang Z., Rank E.}:
\newblock The finite cell method for three-dimensional problems of solid
  mechanics.
\newblock \emph{Computer Methods in Applied Mechanics and Engineering 197}, 45
  (2008), 3768--3782.

\bibitem[Elb23]{Elber2023}
\textsc{Elber G.}:
\newblock A review of a b-spline based volumetric representation: Design,
  analysis and fabrication of porous and/or heterogeneous geometries.
\newblock \emph{Computer-Aided Design 163} (2023), 103587.

\bibitem[ELK97]{Elber97}
\textsc{Elber G., Lee I.-K., Kim M.-S.}:
\newblock Comparing offset curve approximation methods.
\newblock \emph{IEEE Computer Graphics and Applications 17}, 3 (1997), 62--71.

\bibitem[Far08]{FaroukiBook}
\textsc{Farouki R.}:
\newblock \emph{Pythagorean-Hodograph Curves}.
\newblock Springer, Berlin, 2008.

\bibitem[FLLF21]{Feng2021}
\textsc{Feng J., Liu B., Lin Z., Fu J.}:
\newblock Isotropic porous structure design methods based on triply periodic
  minimal surfaces.
\newblock \emph{Materials \& Design 210} (2021), 110050.

\bibitem[GCDL22]{Gao2022}
\textsc{Gao D., Chen J., Dong Z., Lin H.}:
\newblock Connectivity-guaranteed porous synthesis in free form model by
  persistent homology.
\newblock \emph{Computers \& Graphics 106} (2022), 33--44.

\bibitem[GGL24]{Gao2024}
\textsc{Gao D., Gao Y., Lin H.}:
\newblock Periodic implicit representation, design and optimization of porous
  structures using periodic b-splines.
\newblock \emph{arXiv preprint arXiv:2402.12076 171} (2024), 103703.

\bibitem[HE21]{Hong2021}
\textsc{Hong Q.~Y., Elber G.}:
\newblock Conformal microstructure synthesis in trimmed trivariate based
  v-reps.
\newblock \emph{Computer-Aided Design 140} (2021), 103085.

\bibitem[HEK23]{Hong2023}
\textsc{Hong Q.~Y., Elber G., Kim M.-S.}:
\newblock Implicit functionally graded conforming microstructures.
\newblock \emph{Computer-Aided Design 162} (2023), 103548.

\bibitem[HL93]{Hoschek93}
\textsc{Hoschek J., Lasser D.}:
\newblock \emph{Fundamentals of Computer Aided Geometric Design}.
\newblock AK Peters, Wellesley, MA, 1993.

\bibitem[HL21]{Hu2021}
\textsc{Hu C., Lin H.}:
\newblock Heterogeneous porous scaffold generation using trivariate b-spline
  solids and triply periodic minimal surfaces.
\newblock \emph{Graphical Models 115} (2021), 101105.

\bibitem[Hof89]{Hoffmann1989}
\textsc{Hoffmann C.}:
\newblock \emph{Geometric \& Solid Modeling: An Introduction}.
\newblock Morgan Kaufmann, 1989.

\bibitem[HPKE19]{Hong2019}
\textsc{Hong Q.~Y., Park Y., Kim M.-S., Elber G.}:
\newblock Trimming offset surface self-intersections around near-singular
  regions.
\newblock \emph{Computers \& Graphics 82} (2019), 84--94.

\bibitem[HRAB24]{Hirschler2024}
\textsc{Hirschler T., R.Bouclier, Antolin P., Buffa A.}:
\newblock Reduced order modeling based inexact feti-dp solver for lattice
  structures.
\newblock \emph{International Journal for Numerical Methods in Engineering
  e7419} (2024).

\bibitem[Hug00]{Hughes2000}
\textsc{Hughes T.}:
\newblock \emph{The finite element method: linear static and dynamic finite
  element analysis}.
\newblock Dover Publications, Mineola, NY, 2000.

\bibitem[Iri24]{IRIT24}
\textsc{Irit}:
\newblock The irit geometric modeling environment., 2024.
\newblock \href{http://www.cs.technion.ac.il/~gershon/irit}
  {http://www.cs.technion.ac.il/$\sim$gershon/irit}.

\bibitem[Joh]{customquad}
\textsc{Johansson A.}:
\newblock Customquad.
\newblock https://github.com/augustjohansson/ustomquad.

\bibitem[KAH{\etalchar{*}}21a]{Korshunova2021b}
\textsc{Korshunova N., Alaimo G., Hosseini S., Carraturo M., Reali A., Niiranen
  J., Auricchio F., Rank E., Kollmannsberger S.}:
\newblock Bending behavior of octet-truss lattice structures: Modelling
  options, numerical characterization and experimental validation.
\newblock \emph{Materials \& Design 205} (2021), 109693.

\bibitem[KAH{\etalchar{*}}21b]{Korshunova2021a}
\textsc{Korshunova N., Alaimo G., Hosseini S., Carraturo M., Reali A., Niiranen
  J., Auricchio F., Rank E., Kollmannsberger S.}:
\newblock Image-based numerical characterization and experimental validation of
  tensile behavior of octet-truss lattice structures.
\newblock \emph{Additive Manufacturing 41} (2021), 101949.

\bibitem[KHL19]{Kochmann2019}
\textsc{Kochmann D., Hopkins J., L.Valdevit}:
\newblock Multiscale modeling and optimization of the mechanics of hierarchical
  metamaterials.
\newblock \emph{MRS Bulletin 44}, 10 (2019), 773--781.

\bibitem[LAR{\etalchar{*}}21]{Lehder2021}
\textsc{Lehder E., Ashcroft A., R.Wildman, L.Ruiz-Cantu, I.Maskery}:
\newblock A multiscale optimisation method for bone growth scaffolds based on
  triply periodic minimal surfaces.
\newblock \emph{Biomechanics and Modeling in Mechanobiology 20} (2021),
  2085--2096.

\bibitem[LC87]{Lorensen87}
\textsc{Lorensen W.~E., Cline H.~E.}:
\newblock Marching cubes: A high resolution 3d surface construction algorithm.
\newblock \emph{Computer Graphics (SIGGRAPH'87 Proc.) 21}, 4 (1987), 163–169.

\bibitem[LM14]{McMains2014}
\textsc{Li W., McMains S.}:
\newblock A sweep and translate algorithm for computing voxelized 3d minkowski
  sums on the gpu.
\newblock \emph{Computer-Aided Design 46} (2014), 90--100.

\bibitem[Mae99]{Maekawa99}
\textsc{Maekawa T.}:
\newblock An overview of offset curves and surfaces.
\newblock \emph{Computer-Aided Design 31}, 3 (1999), 165--173.

\bibitem[MAE19]{Massarwi2019}
\textsc{Massarwi F., Antolin P., Elber G.}:
\newblock Volumetric untrimming: Precise decomposition of trimmed trivariates
  into tensor products.
\newblock \emph{Computer Aided Geometric Design 71} (2019), 1--15.

\bibitem[ME16]{Massarwi2016}
\textsc{Massarwi F., Elber G.}:
\newblock A b-spline based framework for volumetric object modeling.
\newblock \emph{Computer-Aided Design 78} (2016), 36--47.

\bibitem[MMAE18]{Massarwi2018}
\textsc{Massarwi F., Machchhar J., Antolin P., Elber G.}:
\newblock Hierarchical, random and bifurcation tiling with heterogeneity in
  micro-structures construction via functional composition.
\newblock \emph{Computer-Aided Design 102} (2018), 148--159.

\bibitem[MS18]{main2018}
\textsc{Main A., Scovazzi G.}:
\newblock The shifted boundary method for embedded domain computations. part i:
  Poisson and stokes problems.
\newblock \emph{Journal of Computational Physics 372} (2018), 972--995.

\bibitem[MSA{\etalchar{*}}18]{Maskery2018}
\textsc{Maskery I., Sturm L., Aremu A., Panesar A., Williams C., Tuck C.,
  Wildman R., Ashcroft I., Hague R.}:
\newblock Insights into the mechanical properties of several triply periodic
  minimal surface lattice structures made by polymer additive manufacturing.
\newblock \emph{Polymer 152} (2018), 62--71.

\bibitem[nTo]{nTopology}
\textsc{nTopology:}:.
\newblock https://www.ntop.com.

\bibitem[PFV{\etalchar{*}}11]{Pasko2011}
\textsc{Pasko A., Fryazinov O., Vilbrandt T., Fayolle P.-A., Adzhiev V.}:
\newblock Procedural function-based modelling of volumetric microstructures.
\newblock \emph{Graphical Models 73}, 5 (2011), 165--181.

\bibitem[PK08]{Kobbelt}
\textsc{Pavic D., Kobbelt L.}:
\newblock High-resolution volumetric computation of offset surfaces with
  feature preservation.
\newblock \emph{Computer Graphics Forum 27}, 2 (2008), 165--174.

\bibitem[RR86]{Rossignac1986}
\textsc{Rossignac J.~R., Requicha A.~A.}:
\newblock Offsetting operations in solid modelling.
\newblock \emph{Computer Aided Geometric Design 3}, 2 (1986), 129--148.

\bibitem[Say]{algoim}
\textsc{Saye R.~I.}:
\newblock Algoim, algorithms for implicitly defined geometry, level set
  methods, and voronoi implicit interface methods.
\newblock https://algoim.github.io.

\bibitem[Say14]{saye2014high}
\textsc{Saye R.~I.}:
\newblock High-order methods for computing distances to implicitly defined
  surfaces.
\newblock \emph{Communications in Applied Mathematics and Computational Science
  9}, 1 (2014), 107--141.

\bibitem[Say15]{saye2015high}
\textsc{Saye R.~I.}:
\newblock High-order quadrature methods for implicitly defined surfaces and
  volumes in hyperrectangles.
\newblock \emph{SIAM Journal on Scientific Computing 37}, 2 (2015),
  A993--A1019.

\bibitem[Say22]{saye2022high}
\textsc{Saye R.~I.}:
\newblock High-order quadrature on multi-component domains implicitly defined
  by multivariate polynomials.
\newblock \emph{Journal of Computational Physics 448} (2022), 110720.

\bibitem[SDRW22]{basix}
\textsc{Scroggs M.~W., Dokken J.~S., Richardson C.~N., Wells G.~N.}:
\newblock Construction of arbitrary order finite element degree-of-freedom maps
  on polygonal and polyhedral cell meshes.
\newblock \emph{ACM Transactions on Mathematical Software 48}, 2 (2022),
  18:1--18:23.

\bibitem[Sol]{IntactSolutions}
\textsc{Solutions I.}:
\newblock Immersed method of moments\textsuperscript{TM} (imm).
\newblock
  https://intact-solutions.com/rev/2023/06/immersed-method-of-moments-imm.

\bibitem[WGG99]{Wyvill1999}
\textsc{Wyvill B., Guy A., Galin E.}:
\newblock Extending the csg tree. warping, blending and boolean operations in
  an implicit surface modeling system.
\newblock \emph{Computer Graphics Forum 18}, 2 (1999), 149--158.

\bibitem[WJH{\etalchar{*}}22]{Wang2022}
\textsc{Wang S., Jiang Y., Hu J., Fan X., Luo Z., Liu Y., Liu L.}:
\newblock Efficient representation and optimization of tpms-based porous
  structures for 3d heat dissipation.
\newblock \emph{Computer-Aided Design 142} (2022), 103123.

\bibitem[WM13]{CLWang}
\textsc{Wang C.~C., Manocha D.}:
\newblock Gpu-based offset surface computation using point samples.
\newblock \emph{Computer-Aided Design 45}, 2 (2013), 321--330.

\bibitem[WRC{\etalchar{*}}20]{Wang2020}
\textsc{Wang Y., Ren X., Chen Z., Jiang Y., Cao X., Fang S., Zhao T., Li Y.,
  Fang D.}:
\newblock Numerical and experimental studies on compressive behavior of gyroid
  lattice cylindrical shells.
\newblock \emph{Materials \& Design 186} (2020), 108340.

\bibitem[YL23]{Yan2023}
\textsc{Yan J., Lin H.}:
\newblock Reasonable thickness determination for implicit porous sheet
  structure using persistent homology.
\newblock \emph{Computers \& Graphics 115} (2023), 236--245.

\bibitem[YRL{\etalchar{*}}19]{Yan2019}
\textsc{Yan X., Rao C., Lu L., Sharf A., Zhao H., Chen B.}:
\newblock Strong 3d printing by tpms injection.
\newblock \emph{IEEE Transactions on Visualization and Computer Graphics 26},
  10 (2019), 3037--3050.

\bibitem[ZBQC13]{Zanni2013}
\textsc{Zanni C., Bernhardt A., Quiblier M., Cani M.-P.}:
\newblock Scale-invariant integral surfaces.
\newblock \emph{Computer Graphics Forum 32}, 8 (2013), 219--232.

\bibitem[ZEE23]{Zwar2023}
\textsc{Zwar J., Elber G., Elgeti S.}:
\newblock Shape optimization for temperature regulation in extrusion dies using
  microstructures.
\newblock \emph{The Journal of Mechanical Design 145}, 1 (2023), 012004.

\bibitem[Zha05]{Zhao2005}
\textsc{Zhao H.}:
\newblock A fast sweeping method for eikonal equations.
\newblock \emph{Mathematics of Computation 74}, 250 (2005), 603--627.

\end{thebibliography}

\appendix
\section{Simulation}
\label{sec-appendix}

The finite element unfitted discretization paradigm briefly introduced in Section~\ref{sec-analysis-implicit} allows to decouple the domain
definition from the solution discretization, that is based on a background grid independent from the geometry, what offers great
flexibility.  Nevertheless, this flexibility does not come for free.
Unfitted methods present a series of challenges as, for instance: the
need of imposing essential (Dirichlet) boundary conditions in a weak
sense through Nitsche-like methods, and the possible stability
problems associated with their imposition (see, e.g., \cite{buffa2020});
the possible ill-conditioning of the resulting linear system
associated with the potentially arbitrary small contribution of some cut
basis functions~\cite{deprenter2023}; or the computation of integrals
over cut elements in $\mathcal{G}_{\text{cut}}$, that will be addressed below.

In order to better illustrate the method, we focus on a linear elasticity problem, that is governed by a variational problem of the form:
\begin{equation}
    \int_\Omega \sigma(u):\varepsilon(v)\text{d}x =\int_\Omega f\cdot v\text{d}x + \int_{\Gamma_N} t\cdot v\text{d}x\,,
    \label{eqn-cutfem}
\end{equation}
where $u:\Omega\to\Reals^3$ is the elastic deformation field, and $v:\Omega\to\Reals^3$ the associated test or virtual deformation; $\varepsilon(v) = \nabla^{\text{s}} v$ is the infinitesimal strain tensor and
$\sigma(u)=2\mu\varepsilon(u)+\lambda\text{tr}(\varepsilon(u)) I$ the Cauchy stress tensor, being $\lambda$ and $\mu$ the Lam\'e material parameters; finally, $f:\Omega\to\Reals^3$ is the distributed body load and $t:\Gamma_N\to\Reals^3$ the applied traction (Neumann) boundary condition on a boundary subset $\Gamma_N\subset\partial\Omega$.
In this setting, and for the sake simplicity, we assumed that prescribed displacements (Dirichlet) boundary conditions are only applied on $\Gamma_D\subset\partial\Omega\cap\partial\mathcal{G}$, such that $\Gamma_D=\partial\Omega\setminus\Gamma_N$, and therefore can be enforced strongly.
\begin{remark}
    In the numerical examples presented in Section~\ref{sec-results}, the domain $\Omega$ was defined through implicits in the parametric space $D$ of a macro-shape $\mathcal{T}$ that maps them into the Euclidean space.
    In such a case, the PDE at hand is discretized by means of a Cartesian grid $\mathcal{G}$ covering $D$, and, consequently, the integrals in~\eqref{eqn-cutfem} must be pulled-back from the Euclidean to the parametric space through $\mathcal{T}^{-1}$.
\end{remark}

After discretizing $u$ and $v$ in the active part of the background grid, i.e., in $\mathcal{G}_{\text{in}}\cup\mathcal{G}_{\text{cut}}$,\footnote{The degrees-of-freedom associated with basis functions whose support only intersect $\mathcal{G}_{\text{out}}$ are not considered in the problem.} Equation~\eqref{eqn-cutfem} yields a linear system of equations whose resolution gives the coefficients of the approximated PDE solution.
Nonetheless, in order to assemble the system, the integrals involved must be precisely evaluated inside $\Omega$. Thus, these integrals can be computed element-wise for a generic quantity $\alpha:\Omega\to\Reals$ as:
\begin{equation}
    \int_\Omega \alpha(x)\text{d}x =
    \sum_{Q\in\mathcal{G}_{\text{in}}} \int_Q  \alpha(x)\text{d}x
    +    \sum_{Q\in\mathcal{G}_{\text{cut}}} \int_{Q\cap\Omega} \alpha(x)\text{d}x.
\end{equation}
For the first term, the integrals over $\mathcal{G}_{\text{in}}$ are evaluated in a standard way by means of Gauss-Legendre quadrature rules.
On the other hand, the integrals over the active part of the cut elements, i.e., $\mathcal{G}_{\text{cut}}\cap\Omega$, require special treatment.
By leveraging on the implicit definition of the domain $\Omega$ described in previous sections, we create custom quadrature rules for every single element in $\mathcal{G}_{\text{cut}}$ by using the methods proposed in~\cite{saye2015high,saye2022high} (through their implementation in~\cite{algoim}).
These methods provide highly accurate and efficient quadrature rules when compared with classical alternatives as, for instance, octree based approaches~\cite{duester2008}.
The same algorithms are applied for boundary integrals.

\begin{remark}
    In order to mitigate the possible ill-conditioning of the linear system of equations caused by arbitrary small cut elements, as proposed in~\cite{burman2016}, we add an additional stabilization term to Equation~\eqref{eqn-cutfem}, namely, $\int_{\mathcal{G}_{\textrm{cut}}\setminus\Omega} \sigma(u):\varepsilon(v)\textrm{d}x$.
    As detailed in that work, this stabilization results in a uniform bound on the linear system's condition number for any discretization degree $p\geq 1$, while preserves full order of convergence of the numerical scheme only for $p=1$.
\end{remark}

\end{document}